\documentclass[10pt]{article}
\usepackage[preprint]{tmlr}

\usepackage[utf8]{inputenc}
\usepackage[T1]{fontenc}
\usepackage[hidelinks]{hyperref}
\usepackage{url}
\usepackage{booktabs}
\usepackage{array}
\usepackage{amsfonts}
\usepackage{nicefrac}
\usepackage{microtype}
\usepackage{amsmath}
\usepackage{amssymb}
\usepackage{amsthm}
\usepackage{graphicx}
\graphicspath{{figures/}}
\usepackage{xcolor}
\usepackage{float}
\newfloat{algorithm}{tbp}{loa}
\floatname{algorithm}{Algorithm}
\newenvironment{algorithmic}[1][]{\begin{list}{}{\leftmargin=1.5em\itemsep=2pt\parsep=0pt\topsep=4pt}\ttfamily\small}{\end{list}}
\newcommand{\State}{\item}
\newcommand{\For}[1]{\item \textit{for} #1 \textit{do}}
\newcommand{\EndFor}{\item \textit{end for}}

\usepackage{caption}
\usepackage{subcaption}

\newtheorem{theorem}{Theorem}[section]
\newtheorem{proposition}[theorem]{Proposition}

\theoremstyle{remark}

\theoremstyle{plain}

\newcommand{\KL}{\mathrm{KL}}
\newcommand{\sKL}{D_{\mathrm{sKL}}^{\Lambda}}
\newcommand{\SNR}{\mathrm{SNR}}

\title{Soft-Min-SNR as a Smoothed KL: Interpreting and Matching SNR-Based Loss Weights for Diffusion Training}

\author{\name Lei Li \email lei.li@lunarai.llc \\
        \addr LunarAI LLC, Newark, DE, USA
        \AND
        \name Yuexiao Dong \email yuexiao.dong@temple.edu \\
        \addr Department of Statistics, Operations, and Data Science, Fox School of Business\\
        Temple University, Philadelphia, PA, USA
        \AND
        \name Edoardo M. Airoldi \email airoldi@temple.edu \\
        \addr Department of Statistics, Operations, and Data Science, Fox School of Business\\
        Temple University, Philadelphia, PA, USA}

\begin{document}

\maketitle

\begin{abstract}
Diffusion models are trained on a weighted sum of denoising losses, one per noise level. Popular weightings such as Min-SNR and Soft-Min-SNR were introduced as heuristics, each with a threshold set by tuning. We obtain Soft-Min-SNR from a comparison of distributions. At each timestep, the model predicts an independent noisy replica of the clean sample, and we compare its predictive law with the replica's true law by a Kullback--Leibler (KL) divergence after smoothing both with a Gaussian kernel of variance $\lambda$. The resulting objective weights the squared noise-prediction error by $w(t,\lambda)=\sigma_t^2/(\sigma_t^2+\lambda)$. This weight equals Soft-Min-SNR with threshold $1/\lambda$ in the variance-exploding parameterization and is a constant multiple of Soft-Min-SNR and of P2 with unit exponent under variance-preserving schedules. It agrees with Min-SNR at threshold $1/\lambda$ at the highest SNRs and is smaller elsewhere. Smoothing leaves the optimal noise predictor unchanged, and a larger $\lambda$ lowers the share of the stochastic gradient's second moment that comes from high-SNR timesteps, which a global rescaling of the loss cannot do. On CIFAR-10, a Min-SNR threshold near $1/(2\lambda)$ gives the closest final FID at $\lambda=0.05$, and one between $1/(2\lambda)$ and $1/\lambda$ at $\lambda=0.2$. With the linear schedule, the weight reaches early FID targets with 10 to 21\% fewer updates than uniform weighting, but the lead is gone by mid-training. Final FID shows no consistent advantage over uniform weighting on CIFAR-10 or CelebA-64. Drawing training noise levels from the density that k-diffusion pairs with its default threshold lowers FID by $1.3$ to $1.6$ for both uniform weighting and our weight but leaves the gap between them in place. The construction accounts for the functional form of Soft-Min-SNR and reads its threshold as a cap on the precision of the smoothed comparison.
\end{abstract}
\section{Introduction}
\label{sec:intro}

Training a diffusion model means solving many denoising problems at once. A clean sample $x_0$, drawn from the data distribution $p_0$, is corrupted as $x_t=\sqrt{\bar\alpha_t}\,x_0+\sigma_t\epsilon$, with $\epsilon\sim\mathcal N(0,I)$, $t\in\{1,\dots,T\}$ and $\bar\alpha_t=\prod_{s\le t}(1-\beta_s)$ for a noise schedule $\beta_1,\dots,\beta_T$, and a single network $\epsilon_\theta(x_t,t)$ is asked to recover the noise at every level, from images that are barely perturbed to images that are almost pure noise. Diffusion probabilistic models~\citep{sohl2015deep} and score networks trained on Gaussian-perturbed data~\citep{song2019generative} share this structure, and the two were later unified through stochastic differential equations~\citep{songetal2021scorebasedgenerative}. Unless stated otherwise we work with variance-preserving (VP) schedules, $\sigma_t^2=1-\bar\alpha_t$, for which the signal-to-noise ratio $\SNR(t)=\bar\alpha_t/\sigma_t^2$ decreases in $t$; small $t$ therefore means high SNR. Denoising diffusion probabilistic models~\citep{ho2020denoising} average the squared errors over timesteps with equal weights, $L_{\mathrm{simple}}=\mathbb E\|\epsilon-\epsilon_\theta(x_t,t)\|^2$, and throughout the paper ``DDPM'' refers to training with this loss. Where the distinction matters, capital letters such as $X_0$ and $X_t$ denote random variables and lowercase letters their values.

Equal weighting is a choice, and not an obvious one. The denoising problems are not equally hard, and they compete for the capacity of one network and for the direction of each gradient step. Several reweightings have been proposed in response. Min-SNR-$\gamma$~\citep{hang2023minsnr} clips the SNR at a threshold $\gamma$, after observing that the gradients of different timesteps conflict and slow training down. Soft-Min-SNR~\citep{crowson2024hourglass} replaces the clip by the smooth weight $\gamma'/(\gamma'+\SNR)$. P2~\citep{choi2022perception} uses $1/(k+\SNR)^{\gamma_{\mathrm{P2}}}$ to emphasize the noise levels that matter perceptually, the $v$-parameterization comes with truncated-SNR and SNR$+1$ weightings~\citep{salimans2022progressive}, and VDM, the variational diffusion model of~\citet{kingma2021variational}, weights by $-\tfrac{d}{dt}\log\SNR(t)$. Under uniform-$t$ sampling all of these except the VDM weight are functions of $\SNR(t)$ alone; Table~\ref{tab:weights} lists the ones we compare. A second group of methods leaves the per-sample weight alone and changes how often each noise level is sampled: EDM uses a lognormal proposal~\citep{karras2022elucidating}, \citet{esser2024scaling} a logit-normal density and \citet{hang2025improved} adapted schedules, while \citet{go2023negative} and \citet{karras2024analyzing} learn the weights during training. A sampling density and a per-sample multiplier combine into one effective weight per unit log-SNR~\citep{kingma2023understanding}, so the two groups can be put on a common scale (Appendix~\ref{app:edm}).

Each of these weights was introduced for its own reasons, and each comes with a threshold that is tuned by hand. Some basic questions are left open. Why does Soft-Min-SNR have the form $\gamma'/(\gamma'+\SNR)$ rather than some other smooth clip? What does its threshold $\gamma'$ measure? And if a practitioner has tuned $\gamma$ for Min-SNR, or $k$ for P2, which Soft-Min-SNR threshold corresponds to it? We answer them by obtaining Soft-Min-SNR from a comparison of predictive distributions after Gaussian smoothing.

There is a body of work that explains what weighted diffusion losses estimate. \citet{song2021maximum} showed that a particular weighting turns the score-matching loss into a bound on the negative log-likelihood. \citet{kim2022soft} extended the bound to weights that increase with diffusion time relative to this likelihood weighting, as an average of KL divergences between perturbed data and model over a random diffusion time whose distribution function is the normalized relative weight. \citet{kingma2023understanding} wrote every weighted diffusion loss as a weighted integral of evidence lower bounds (ELBOs) over noise levels; when the weight is monotone in log-SNR, the loss is the ELBO of data augmented with Gaussian noise at a random level whose distribution the weight determines. \citet{shi2025demystifying} read reweighted losses as time-dependent variational bounds. These results tell us how to interpret a given weight, but they do not single out the soft-clip form or connect its threshold to those of other weights. Kingma and Gao in fact propose a sigmoid weighting that, as a function of log-SNR, has the Soft-Min-SNR shape (Section~\ref{sec:relations}). They select it by experiment, guided by their monotone-weighting result, and tune its offset by search; their Appendix~I relates monotone weightings to low-bit training, and \citet{hoogeboom2025simpler} draw the same connection for the sigmoid. Through their theorem the sigmoid corresponds to augmentation levels spread logistically around its offset, whereas we obtain the curve from a single smoothing level. Their monotone-weighting result covers weights that are monotone per unit log-SNR; under uniform-$t$ training on the linear schedule the implied weight is not, for DDPM or for our loss (Figure~\ref{fig:weights}b).

The tool we use is older than diffusion models. When two distributions are convolved with the same Gaussian kernel before they are compared, the comparison becomes coarser, and divergences that are uninformative for the original pair can become informative. Adding the same noise to both distributions is what makes the Jensen--Shannon divergence useful for distributions with nearly disjoint supports~\citep{arjovsky2017towards}. \citet{zhang2018spread} turned the idea into a divergence, the spread divergence, of which the KL divergence between Gaussian-smoothed distributions used here is a special case. Kernel smoothing has also been used closer to diffusion: smoothing the empirical score mitigates memorization~\citep{gabriel2025kernelsmoothed}, generative drifting has been related to score matching on smoothed distributions~\citep{turan2026drifting}, and smoothing underlies regularized kernel KL divergences~\citep{chazal2024regularized} and structure-aware divergences that replace $p$ by $Zp$ for a similarity operator $Z$~\citep{sahasrabuddhe2026structure}. None of this work uses smoothing to obtain a per-timestep training weight, which is the step taken here.

Our starting point is a simple thought experiment. Suppose that, alongside $x_t$, we had an independent noisy replica of $x_0$ at the same noise level. Given $x_0$, the replica is Gaussian with mean $\sqrt{\bar\alpha_t}\,x_0$ and covariance $\sigma_t^2I$, and the network can predict it by passing its own estimate of $x_0$ through the same Gaussian channel. The two laws then share a covariance, and the KL divergence between them is exactly half the DDPM loss at that timestep. At this stage nothing has changed: the replica is only a way of reading the loss. The change comes when both laws are smoothed with a Gaussian kernel of variance $\lambda$ before they are compared. Both covariances become $(\sigma_t^2+\lambda)I$, and the expected smoothed divergence is half of
\begin{equation}
\mathcal{L}_\lambda(\theta)
=
\mathbb{E}_{t,x_0,\epsilon}
\left[
w(t,\lambda)
\left\|
\epsilon-\epsilon_\theta(x_t,t)
\right\|^2
\right],
\qquad
w(t,\lambda)
=
\frac{\sigma_t^2}{\sigma_t^2+\lambda},
\label{eq:intro_loss}
\end{equation}
where $t$ is uniform on $\{1,\dots,T\}$ and $\lambda\ge0$ is the smoothing variance. Setting $\lambda=0$ recovers DDPM. For $\lambda>0$ the loss is damped at timesteps whose noise variance $\sigma_t^2$ is small compared with $\lambda$, that is, at high SNR, and left nearly unchanged elsewhere.

In the variance-exploding (VE) parameterization, where \citet{crowson2024hourglass} defined Soft-Min-SNR, the weight in Eq.~\ref{eq:intro_loss} is Soft-Min-SNR with threshold $\gamma'=1/\lambda$, and under VP schedules it is a constant multiple of Soft-Min-SNR. The replica picture shows why. The extra noise adds to the replica's variance, so the precision with which the predicted and true means are compared can never exceed $1/\lambda$. At low SNR this cap is irrelevant; at high SNR it is what limits the weight, and it plays the role of the Soft-Min-SNR threshold. The same algebra places the weight in the P2 family with unit exponent~\citep{choi2022perception}, and Kingma and Gao's sigmoid weighting is the same curve when log-SNR is sampled uniformly. Min-SNR behaves differently. Its hard clip agrees with our weight at $\gamma=1/\lambda$ only to leading order at high SNR, and the two separate in the transition band, which is where the weight changes fastest.

Two further results describe what smoothing leaves alone and what it moves. The expected smoothed divergence splits into a conditional mutual information, which does not involve the network, and a predictive KL divergence, which is minimized by the posterior mean of the clean sample (Proposition~\ref{prop:predictive}). Over unrestricted functions the optimal noise predictor is therefore $\mathbb E[\epsilon\mid X_t]$ whatever the value of $\lambda$; in practice one network of limited capacity serves all timesteps, and the weight decides where its approximation errors go. The second result concerns optimization: a larger $\lambda$ lowers the share of the single-sample gradient's second moment that comes from any fixed set of highest-SNR timesteps, for any per-timestep second moments (Proposition~\ref{prop:share}).

We test these relations on CIFAR-10 and CelebA-64 at a fixed compute budget, comparing the smoothed-KL weight with uniform weighting and with Min-SNR. The clearest result concerns the Min-SNR threshold. At $\lambda=0.05$ on CIFAR-10, Min-SNR with $\gamma$ near $1/(2\lambda)$ matches our weight in final Fr\'echet Inception Distance (FID) more closely than the high-SNR match $\gamma=1/\lambda$; at $\lambda=0.2$ the matching threshold lies between the two, near $0.7/\lambda$, and at the other values of $\lambda$ the data do not separate them (Section~\ref{sec:matching}). Against uniform weighting the picture is mixed. With the linear schedule on CIFAR-10 our weight reaches early FID targets sooner than DDPM, but the lead is gone by mid-training (Section~\ref{sec:speed}), and final FID shows no consistent advantage on either dataset. The thresholds used in practice transfer poorly: the default $\gamma'=1/\sigma_{\mathrm{data}}^2$ falls between two settings that are $1.3$ and $1.6$ FID worse than uniform weighting here. Training with the noise-level density that k-diffusion pairs with that default lowers FID for both weights but leaves the gap in place, so the density does not explain it (Section~\ref{sec:density}).

The rest of the paper is organized as follows. Section~\ref{sec:method} develops the smoothed-KL construction and relates the weight to Soft-Min-SNR, P2 and Min-SNR. Section~\ref{sec:theory} examines how the weight changes the stochastic gradient. Section~\ref{sec:experiments} reports the experiments on CIFAR-10 and CelebA-64, and Section~\ref{sec:discussion} discusses the results and their limitations. Proofs and additional analyses are in the appendices.

\section{Method}
\label{sec:method}

This section turns the thought experiment of the introduction into a derivation. We first define the smoothed KL for general distributions and record two properties that make it a reasonable training criterion. Section~\ref{sec:ddpm} applies it to the replica at each DDPM timestep, and Section~\ref{sec:relations} compares the resulting weight with the heuristics in Table~\ref{tab:weights}. Proofs of all results are in Appendix~\ref{sec:proofs}.

\begin{table}[t]
\centering
\footnotesize
\setlength{\tabcolsep}{4pt}
\renewcommand{\arraystretch}{1.15}
\caption{Per-sample weights on the squared noise-prediction error under uniform timestep sampling, with $s=\SNR(t)$. Under VP schedules our weight is a constant multiple of Soft-Min-SNR and of P2 with unit exponent, and at high SNR it agrees with Min-SNR at $\gamma=1/\lambda$ (Section~\ref{sec:relations}).}
\label{tab:weights}
\begin{tabular}{@{}lll@{}}
\toprule
Method & Weight $w(t)$ & Interpretation \\
\midrule
DDPM & $1$ & Uniform \\
Min-SNR & $\min(s,\gamma)/s$ & Hard clipping \\
Soft-Min-SNR & $\gamma'/(\gamma'+s)$ & Smooth clipping \\
P2 & $(k+s)^{-\gamma_{\mathrm{P2}}}$ & Perceptual \\
Ours & $\sigma_t^2/(\sigma_t^2+\lambda)$ & Smoothed KL \\
\bottomrule
\end{tabular}
\end{table}

\begin{figure}[t]
\centering
\includegraphics[width=\linewidth]{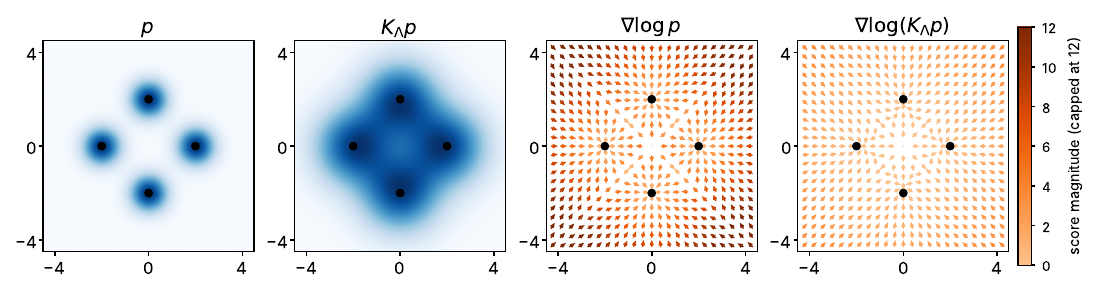}
\caption{The smoothing operator $K_\Lambda$ on a four-mode Gaussian mixture (modes at $(\pm2,0)$ and $(0,\pm2)$, component variance $0.3$, $\Lambda=I$). From left: density, smoothed density, score field and smoothed score field. Arrows show direction and color shows magnitude on a common scale. Within the plotted square the smoothed score has magnitude at most $3.8$, against up to $16.5$ for the original score (capped at $12$ in the color scale), and its direction changes more gently between modes. All fields are computed analytically.}
\label{fig:toy_scores}
\end{figure}

We begin with the comparison itself, for arbitrary distributions on $\mathbb{R}^d$. For a positive semidefinite $d\times d$ matrix $\Lambda$, the smoothing operator $K_\Lambda$ maps a distribution $p$ to the law of $U+V$, where $U\sim p$ and $V\sim\mathcal N(0,\Lambda)$ are independent, so that $K_0p=p$; when $\Lambda$ is positive definite, $K_\Lambda p$ has density $(K_\Lambda p)(x)=\int\mathcal N(x-y;\,0,\Lambda)\,p(dy)$, with $\mathcal N(\cdot\,;\mu,\Sigma)$ the $\mathcal N(\mu,\Sigma)$ density. The smoothed KL divergence is
\begin{equation}
\sKL(q \,\|\, p) = \KL(K_\Lambda q \,\|\, K_\Lambda p),
\label{eq:smoothed_kl}
\end{equation}
where $\KL(q\|p)=\int\log(dq/dp)\,dq$ if $q$ is absolutely continuous with respect to $p$ and $+\infty$ otherwise.
Figure~\ref{fig:toy_scores} shows what the kernel does to a four-mode Gaussian mixture. Smoothing spreads mass between nearby modes, and within the plotted region the smoothed score varies more gently between modes and is smaller in magnitude than the original score away from them. Theorem~\ref{thm:score} in Appendix~\ref{app:tweedie} explains the change: the smoothed score is the posterior average of the original score under a Gaussian channel. Smoothing does not bound scores in general. For $p=\mathcal N(0,vI)$, $\nabla\log(K_{\lambda I}p)(x)=-x/(v+\lambda)$, which grows without bound in $\|x\|$, and the tail scores of the mixture in Figure~\ref{fig:toy_scores} grow linearly in the same way.

For $\sKL$ to serve as a training criterion it should be a weaker comparison than KL that becomes KL as the kernel shrinks. The two propositions below show that it does both.

\begin{proposition}[Contraction]
\label{prop:contraction}
For any distributions $q,p$ on $\mathbb{R}^d$ and any positive semidefinite $\Lambda$, $\sKL(q\|p) \le \KL(q\|p)$, with equality at $\Lambda=0$.
\end{proposition}

\begin{proposition}[Monotonicity and recovery]
\label{prop:recovery}
For any distributions $q,p$ on $\mathbb{R}^d$,
the smoothed KL divergence is nonincreasing in
the smoothing covariance: if
$0\preceq\Lambda_1\preceq\Lambda_2$
(i.e., $\Lambda_2-\Lambda_1$ is positive
semidefinite), then
$D_{\mathrm{sKL}}^{\Lambda_2}(q\|p)
\le
D_{\mathrm{sKL}}^{\Lambda_1}(q\|p).$
Moreover, $D_{\mathrm{sKL}}^\Lambda(q\|p)
\longrightarrow \KL(q\|p)$ as $\Lambda\to0$, where both sides may be $+\infty$.
\end{proposition}

Proposition~\ref{prop:contraction} is the data-processing inequality. Proposition~\ref{prop:recovery} needs no regularity conditions: monotonicity follows from the semigroup property $K_{\Lambda_2}=K_{\Lambda_2-\Lambda_1}K_{\Lambda_1}$ for $\Lambda_1\preceq\Lambda_2$, and the limit from lower semicontinuity of KL under weak convergence (Appendix~\ref{sec:proofs}). Together they place standard KL at the end of a one-parameter family that decreases as the kernel widens, so $\Lambda$ sets how coarsely the two distributions are compared. What makes the construction useful for diffusion is Gaussian closure, $K_\Lambda\mathcal N(\mu,\Sigma)=\mathcal N(\mu,\Sigma+\Lambda)$: for Gaussian $q=\mathcal N(\mu_q,\Sigma_q)$ and $p=\mathcal N(\mu_p,\Sigma_p)$ with $\Sigma_q+\Lambda$ and $\Sigma_p+\Lambda$ positive definite, $\sKL$ is the usual Gaussian KL with every covariance shifted by $\Lambda$,
\begin{equation}
\sKL(q\|p) = \tfrac{1}{2}\Bigl[\mathrm{tr}\bigl((\Sigma_p+\Lambda)^{-1}(\Sigma_q+\Lambda)\bigr) - d + \delta^\top(\Sigma_p+\Lambda)^{-1}\delta + \log\tfrac{|\Sigma_p+\Lambda|}{|\Sigma_q+\Lambda|}\Bigr],\qquad \delta=\mu_q-\mu_p .
\label{eq:closure}
\end{equation}
Here $|\cdot|$ denotes the determinant. Smoothing thus replaces the information metric $\Sigma_p^{-1}$ of the mean by $(\Sigma_p+\Lambda)^{-1}$~\citep{amari2016information}, and through denoising score matching~\citep{vincent2011connection} it corresponds to comparing scores at a larger effective noise level (Appendix~\ref{app:tweedie}).

\subsection{Application to DDPM}
\label{sec:ddpm}

At each timestep we construct a pair of Gaussian laws whose unsmoothed KL is the simplified DDPM loss, so that Eq.~\ref{eq:closure} turns it into a weighted loss. Fix $t$ and a training pair $(X_0,X_t)$ with $X_t=\sqrt{\bar\alpha_t}\,X_0+\sigma_t\epsilon$. Alongside $X_t$, consider an independent replica of the clean sample at the same noise level,
\begin{equation}
Z_t=\sqrt{\bar\alpha_t}\,X_0+\sigma_t\eta,\qquad \eta\sim\mathcal N(0,I),
\label{eq:replica}
\end{equation}
with $\eta$ independent of $(X_0,\epsilon)$, so that $Z_t$ and $X_t$ are conditionally independent given $X_0$. Given $X_0=x_0$ the replica has the exact law $q_t(\cdot\mid x_0)=\mathcal N(\sqrt{\bar\alpha_t}\,x_0,\,\sigma_t^2I)$, whatever the data distribution. The model predicts the replica from $X_t=x_t$ by passing its estimate of the clean sample, $\hat x_0(x_t,t)=\bigl(x_t-\sigma_t\epsilon_\theta(x_t,t)\bigr)/\sqrt{\bar\alpha_t}$, through the same channel: $p_{\theta,t}(\cdot\mid x_t)=\mathcal N(\sqrt{\bar\alpha_t}\,\hat x_0(x_t,t),\,\sigma_t^2I)$. We write $q_t$ and $p_{\theta,t}$ when the conditioning is clear. Giving the model's law the same covariance $\sigma_t^2I$ as the true law given $X_0$ is the modeling choice of the construction. It makes the KL a pure squared error in the mean, and the weight below is its consequence. The two means differ by $\delta=\sqrt{\bar\alpha_t}(x_0-\hat x_0)=\sigma_t(\epsilon_\theta-\epsilon)$, so without smoothing $\KL(q_t\|p_{\theta,t})=\|\delta\|^2/(2\sigma_t^2)=\frac12\|\epsilon-\epsilon_\theta(x_t,t)\|^2$: predicting the replica reproduces $L_{\mathrm{simple}}$ up to the factor $\frac12$.

Smoothing has an equally concrete meaning here. With $\Lambda=\lambda I$, $K_\Lambda q_t$ and $K_\Lambda p_{\theta,t}$ are the true and the model laws of the replica observed through additional measurement noise,
\begin{equation}
Y_t=Z_t+\sqrt\lambda\,\xi,\qquad \xi\sim\mathcal N(0,I),
\label{eq:measurement}
\end{equation}
with $\xi$ independent of everything else. Equation~\ref{eq:closure} replaces $\sigma_t^2$ by $\sigma_t^2+\lambda$ in the denominator, which gives
\begin{equation}
\sKL(q_t\|p_{\theta,t}) = w(t,\lambda)\,\KL(q_t\|p_{\theta,t}),\qquad w(t,\lambda)=\frac{\sigma_t^2}{\sigma_t^2+\lambda},
\label{eq:loss_marginal}
\end{equation}
and the per-sample training loss $L_t^\lambda=w(t,\lambda)\,\|\epsilon-\epsilon_\theta(x_t,t)\|^2=2\,\sKL(q_t\|p_{\theta,t})$ of Eq.~\ref{eq:intro_loss} (pseudocode in Appendix~\ref{app:pseudocode}). In this reading $\lambda$ is the variance of the measurement noise, and the parameter maps below follow from it. 

Equation~\ref{eq:loss_marginal} holds sample by sample and shows how smoothing changes the loss at a given timestep. Whether it also changes the optimal prediction is a question about the expected loss. In expectation, the loss splits into the uncertainty in the replica, which no predictor can remove, and the error due to the model. Write $q_{\lambda,t}(\cdot\mid x_0)=\mathcal N(\sqrt{\bar\alpha_t}\,x_0,(\sigma_t^2+\lambda)I)$ for the law of $Y_t$ given $X_0=x_0$, $p_{\theta,\lambda,t}(\cdot\mid x_t)=\mathcal N(\sqrt{\bar\alpha_t}\,\hat x_0(x_t,t),(\sigma_t^2+\lambda)I)$ for the model's predictive law given $X_t=x_t$, and $q_{\lambda,t}(\cdot\mid x_t)$ for the true conditional law of $Y_t$ given $X_t=x_t$.

\begin{proposition}[Predictive decomposition]
\label{prop:predictive}
Fix $t$ and $\lambda\ge0$, and suppose $\mathbb E\|X_0\|^2<\infty$ and $\mathbb E\|\hat x_0(X_t,t)\|^2<\infty$. Then
\begin{equation}
\mathbb E\,\KL\bigl(q_{\lambda,t}(\cdot\mid X_0)\,\big\|\,p_{\theta,\lambda,t}(\cdot\mid X_t)\bigr)
= I(X_0;Y_t\mid X_t)+\mathbb E\,\KL\bigl(q_{\lambda,t}(\cdot\mid X_t)\,\big\|\,p_{\theta,\lambda,t}(\cdot\mid X_t)\bigr),
\label{eq:predictive}
\end{equation}
where $I(X_0;Y_t\mid X_t)=\mathbb{E}\left[\log\frac{q_{\lambda,t}(Y_t\mid X_0)}{q_{\lambda,t}(Y_t\mid X_t)}\right]$ is the conditional mutual information between $X_0$ and $Y_t$ given $X_t$ (here $Y_t$ and $X_t$ are conditionally independent given $X_0$). It does not depend on $\theta$ and is nonincreasing in $\lambda$. The left side equals $\frac12\,\mathbb E\,L_t^\lambda$, with $L_t^\lambda=w(t,\lambda)\|\epsilon-\epsilon_\theta(X_t,t)\|^2$. Over all measurable $\hat x_0$, both sides are minimized by $\hat x_0(x_t,t)=\mathbb E[X_0\mid X_t=x_t]$, equivalently $\epsilon_\theta(x_t,t)=\mathbb E[\epsilon\mid X_t=x_t]$, for every $\lambda\ge0$.
\end{proposition}

Proposition~\ref{prop:predictive} fixes the normalization: the expected weighted loss at timestep $t$ is twice the expected predictive KL plus $2I(X_0;Y_t\mid X_t)$. The mutual information does not involve $\theta$ but falls as $\lambda$ grows, so loss values at different $\lambda$ are not comparable. The minimizer over unrestricted predictors is the posterior mean for every $\lambda$, as it is for any positive timestep weight, so smoothing changes how errors at different noise levels are priced, not the population target. A network of limited capacity may not represent that target exactly at every timestep, and the compromise it reaches, through its architecture and through stochastic optimization (Section~\ref{sec:theory}), can depend on $\lambda$. The same logic underlies Noise2Noise~\citep{lehtinen2018noise2noise}, where an independent noisy realization of the target leaves the population solution unchanged; here the replica serves only to read the loss. The decomposition also explains the plug-in covariance. Given $X_t$, the replica has covariance $\sigma_t^2I+\bar\alpha_t\,\mathrm{Cov}(X_0\mid X_t)$, and the model's $\sigma_t^2I$ treats $\hat x_0$ as if it were $x_0$. This affects only the second term of Eq.~\ref{eq:predictive}, and it keeps the weight a function of $t$ rather than of the data.

The reference law has the form of the forward conditional $q(x_t\mid x_0)$, where $q$ denotes the forward diffusion process, but the pair differs from the one compared by the canonical ELBO: the forward posterior $q(x_{t-1}\mid x_t,x_0)$ and the model's reverse kernel $p_\theta(x_{t-1}\mid x_t)$. With the model's reverse variance matched to $\tilde\beta_t=\beta_t(1-\bar\alpha_{t-1})/(1-\bar\alpha_t)$, smoothing those kernels and dropping the ELBO prefactors as $L_{\mathrm{simple}}$ does gives the weight $\tilde\beta_t/(\tilde\beta_t+\lambda)$. Its timestep dependence is different: $\tilde\beta_t$ depends on the step from $t-1$ to $t$, so as a function of $\SNR(t)$ this weight changes with the schedule and the number of steps. Ours is the same function of $\SNR(t)$ for every VP schedule (Eq.~\ref{eq:snr_form}).

Since $\SNR(t)=\bar\alpha_t/\sigma_t^2$, the weight can also be written
\begin{equation}
w(t,\lambda) = \frac{1}{1+\lambda/\sigma_t^2} = \frac{1}{1+\lambda\,\SNR(t)/\bar\alpha_t}.
\label{eq:snr_form}
\end{equation}
The weight is small when $\sigma_t^2\ll\lambda$, at the high-SNR timesteps, and close to $1$ when $\sigma_t^2\gg\lambda$. The replica construction explains this shape. Measured on the clean-sample error $\|x_0-\hat x_0\|^2$, the unsmoothed KL has precision $\bar\alpha_t/\sigma_t^2=\SNR(t)$, which sets how strongly errors are penalized and grows without bound as $\sigma_t^2\to0$. Smoothing adds $\lambda$ to the replica's variance, and the precision falls to
\begin{equation}
\frac{\bar\alpha_t}{\sigma_t^2+\lambda}
=
\left(
\frac{1}{\SNR(t)}
+
\frac{\lambda}{\bar\alpha_t}
\right)^{-1}
\le
\frac{\bar\alpha_t}{\lambda}
\le
\frac{1}{\lambda}.
\label{eq:precision}
\end{equation}
The smoothed precision is a harmonic combination of $\SNR(t)$ and $\bar\alpha_t/\lambda$: it is close to the smaller of the two and moves smoothly between them. Gaussian smoothing therefore caps the precision of the comparison at $1/\lambda$, and does so softly, which is how Soft-Min-SNR clips the SNR. To give a sense of scale, at $\lambda=0.01$ on the linear schedule the cap matters only where $\SNR(t)$ exceeds about $100$, which is the first 27 of the 1000 timesteps.

\subsection{Relation to Soft-Min-SNR, P2 and Min-SNR}
\label{sec:relations}

The precision cap of Eq.~\ref{eq:precision} has a direct consequence for the existing weights. Under VP schedules the smoothed-KL weight is Soft-Min-SNR up to a constant factor.

\begin{proposition}[Exact equivalence with Soft-Min-SNR]
\label{prop:softminsnr}
Let $w_{\mathrm{SMS}}(t,\gamma')=\gamma'/(\gamma'+\SNR(t))$. For any VP schedule ($\bar\alpha_t+\sigma_t^2=1$) and any $\lambda>0$, setting $\gamma'=(1+\lambda)/\lambda$ gives $w(t,\lambda) = \frac{1}{1+\lambda}\, w_{\mathrm{SMS}}(t,\gamma').$
\end{proposition}

The factor $1/(1+\lambda)$ does not depend on $t$: under SGD it is a learning-rate change, and under Adam it cancels apart from the $\varepsilon$ term and, when gradients are clipped, the clipping threshold. \citet{crowson2024hourglass} define Soft-Min-SNR in the variance-exploding parameterization $x_\sigma=x_0+\sigma\epsilon$, for which $\SNR=1/\sigma^2$, as the weight $1/(\sigma^2+1/\gamma')$ on the $x_0$-loss, which is $\sigma^2/(\sigma^2+1/\gamma')$ on the $\epsilon$-loss. Applying the construction of Section~\ref{sec:ddpm} to the VE replica laws $\mathcal N(x_0,\sigma^2I)$ gives $\sigma^2/(\sigma^2+\lambda)$, so there the two weights coincide exactly with $\lambda=1/\gamma'$, and Eq.~\ref{eq:precision} becomes $(1/\SNR+1/\gamma')^{-1}$; their setting $\gamma'=4$ corresponds to $\lambda=0.25$. Under VP schedules a threshold $\gamma'>1$ maps to $\lambda=1/(\gamma'-1)$; no positive $\lambda$ corresponds to $\gamma'\le1$. In both parameterizations, published uses of Soft-Min-SNR are uses of the smoothed-KL weight.

The same curve has a second reading. If the data were Gaussian, $X_0\sim\mathcal N(0,s^2I)$, then in VE $\mathrm{Var}(X_0\mid X_\sigma=x_\sigma)/s^2=\sigma^2/(\sigma^2+s^2)$: our weight with $\lambda=s^2$ is the posterior variance of the clean sample relative to its prior variance, and it damps the timesteps at which $x_\sigma$ already determines $x_0$. The k-diffusion implementation of Soft-Min-SNR sets $\gamma'=1/\sigma_{\mathrm{data}}^2$~\citep{crowson2022kdiffusion}, which is this choice with $s=\sigma_{\mathrm{data}}$. Under VP schedules the Gaussian posterior gives $\gamma'=1/s^2$, smaller than the threshold of Proposition~\ref{prop:softminsnr} by the factor $1+\lambda$, because the prior smooths at the scale of $x_0$ and the replica at the scale of $x_t$. That reading is exact only for Gaussian data and ties $\lambda$ to the data variance; Proposition~\ref{prop:softminsnr} holds for any data and leaves $\lambda$ free.

The same algebra places the weight in the P2 family of~\citet{choi2022perception}. Under VP schedules, with $s=\SNR(t)$ and $\lambda>0$,
\begin{equation}
w(t,\lambda)=\frac{1}{1+\lambda+\lambda s}=\frac1\lambda\cdot\frac{1}{s+(1+\lambda)/\lambda},
\label{eq:p2}
\end{equation}
so up to the constant $1/\lambda$ our weight is P2 with $\gamma_{\mathrm{P2}}=1$ and $k=(1+\lambda)/\lambda$; likewise Soft-Min-SNR with threshold $\gamma'$ is, up to the constant $\gamma'$, P2 with $\gamma_{\mathrm{P2}}=1$ and $k=\gamma'$. Our P2 baseline uses the default $k=1$, $\gamma_{\mathrm{P2}}=1$, the limit of $\lambda\,w(t,\lambda)$ as $\lambda\to\infty$, so it tests one member of the family.

\begin{figure}[t]
\centering
\includegraphics[width=\linewidth]{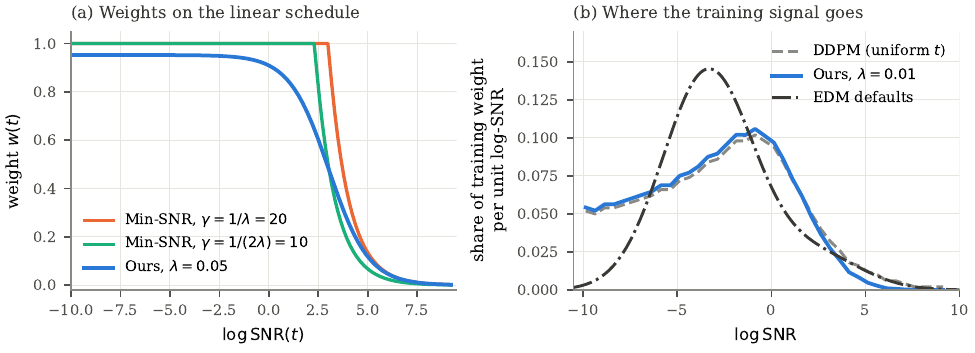}
\caption{(a) Per-sample multipliers against $\log\SNR(t)$ on the linear schedule ($T=1000$). Our weight with $\lambda=0.05$ is everywhere at or below Min-SNR at $\gamma=1/\lambda=20$ and meets it only at the highest SNRs (Proposition~\ref{prop:asymptotic_minsnr}); its low-SNR level is $1/(1+\lambda)$. (b) Effective weight per unit log-SNR, normalized to unit total: the density of $\log\SNR$ induced by the sampling scheme times the per-sample multiplier on the $\epsilon$-loss. For DDPM and our loss the density comes from uniform $t$; for EDM it is the lognormal density over $\sigma$ and the multiplier is $1+1/(\sigma_{\mathrm{data}}^2\SNR)$, with $P_{\mathrm{mean}}=-1.2$, $P_{\mathrm{std}}=1.2$, $\sigma_{\mathrm{data}}=0.5$.}
\label{fig:weights}
\end{figure}

As a function of log-SNR the same curve is a logistic sigmoid, $\gamma'/(\gamma'+\SNR)=\mathrm{sigmoid}(\log\gamma'-\log\SNR)$, where $\mathrm{sigmoid}(z)=1/(1+e^{-z})$ and $\log$ is the natural logarithm. This is the shape of the sigmoid weighting $\mathrm{sigmoid}(b-\log\SNR)$ that~\citet{kingma2023understanding} apply per unit log-SNR, with offset $b$ (their $k$). With log-SNR sampled uniformly their weighting is therefore Soft-Min-SNR with $\gamma'=e^b$, that is, $\lambda=e^{-b}$ (VE) or $\lambda=1/(e^b-1)$ (VP); otherwise the two differ by the density of log-SNR (Figure~\ref{fig:weights}b). Their Theorem~1 reads a monotone weighting as an ELBO on noise-augmented data with a random augmentation level, and for the sigmoid that level is spread out: $\mathrm{sigmoid}(b-\log\SNR)$ is the probability that a logistic variable with location $b$ and unit scale exceeds $\log\SNR$, so the log-SNR of the augmentation is logistic around $b$. Here the curve comes from a single smoothing level, and $b$ is the logarithm of the precision cap in Eq.~\ref{eq:precision}, $b=-\log\lambda$ in VE.

Min-SNR, the hard clip, is more widely used than Soft-Min-SNR, and its threshold $\gamma$ is what practitioners tune. Both weights act mainly at the highest SNRs, so that is where we compare them first.

\begin{proposition}[Comparison with Min-SNR]
\label{prop:asymptotic_minsnr}
Let $\lambda>0$, set $\gamma=1/\lambda$, and let $w_{\mathrm{MinSNR}}(t,\gamma)=\min(\SNR(t),\gamma)/\SNR(t)$. For any schedule with $\bar\alpha_t\in(0,1]$,
\[
w(t,\lambda)\le\frac{1}{1+\lambda\SNR(t)}\le w_{\mathrm{MinSNR}}(t,\gamma).
\]
Moreover, under a VP schedule, whenever $\SNR(t)>\gamma$,
\[
0\le1-\frac{w(t,\lambda)}{w_{\mathrm{MinSNR}}(t,\gamma)}=\frac{\sigma_t^2/\lambda+\sigma_t^2}{1+\sigma_t^2/\lambda}\le\frac{\sigma_t^2}{\lambda}+\sigma_t^2 .
\]
Consequently, $w(t,\lambda)/w_{\mathrm{MinSNR}}(t,\gamma)\to1$ as $\sigma_t^2\to0$.
\end{proposition}
Both weights decay like $1/\SNR(t)$ at high SNR, as does P2 with $\gamma_{\mathrm{P2}}=1$ up to a constant factor; P2 in general decays like $\SNR(t)^{-\gamma_{\mathrm{P2}}}$. In the transition band the picture changes: at $\SNR(t)=1/\lambda$, Min-SNR assigns weight $1$ and our weight at most $1/2$.

The gap suggests that matching the weights in that band may need a Min-SNR threshold below $1/\lambda$. At the reference SNR $s_0=1/\lambda$, our weight is $1/(2+\lambda)$ by Eq.~\ref{eq:p2}, and the clipped branch $\gamma/s_0$ of Min-SNR takes the same value when
\begin{equation}
\gamma_{\mathrm{local}}
=
\frac{s_0}{2+\lambda}
=
\frac{1}{\lambda(2+\lambda)}
\approx
\frac{1}{2\lambda},
\label{eq:local_match}
\end{equation}
which is $9.76$ for $\lambda=0.05$. The answer depends on the reference point. Matching at $s_0=c/\lambda$ gives $\gamma\lambda=c/(1+\lambda+c)$, which approaches $1$ as $c$ grows, consistent with the high-SNR agreement in Proposition~\ref{prop:asymptotic_minsnr}. Matching average weights under uniform timestep sampling gives smaller thresholds still, with $\gamma\lambda$ between $0.06$ and $0.14$ on the linear schedule (Appendix~\ref{app:minsnr_proof}).

Figure~\ref{fig:weights}a shows the weights for $\lambda=0.05$: ours lies below Min-SNR at $\gamma=20$ at every timestep and, in the transition band, closer to $\gamma=10$. Which of these thresholds, if any, gives the same FID after training is an empirical question, and Section~\ref{sec:matching} takes it up. Figure~\ref{fig:weights}b compares effective weights per unit log-SNR and adds EDM~\citep{karras2022elucidating}, whose weighting combines a lognormal sampling density over noise levels with a per-sample multiplier (Appendix~\ref{app:edm}).

\section{What the Reweighting Changes}
\label{sec:theory}

Why would a reweighting help optimization at all? The usual explanations rest on two elementary properties of the weighted quadratic loss, which turn out not to be specific to timestep-dependent weights. Let $\ell_t(\theta)=\|\epsilon-\epsilon_\theta(x_t,t)\|^2$. First, as a function of the predicted mean $\mu_\theta=\sqrt{\bar\alpha_t}\,\hat x_0(x_t,t)$, the auxiliary Gaussian KL of Section~\ref{sec:ddpm} has Hessian $(\Sigma_t+\Lambda)^{-1}$ with $\Sigma_t=\sigma_t^2I$, so its gradient Lipschitz constant is $L_\Lambda=\|(\Sigma_t+\Lambda)^{-1}\|_2\le\|\Sigma_t^{-1}\|_2=L$, with $\|\cdot\|_2$ the spectral norm, and $L_\Lambda/L=w(t,\lambda)$ for $\Lambda=\lambda I$; in the network parameters, curvature also involves the network's derivatives. Second, a deterministic weight $w_t$ multiplies the gradient of $\ell_t$ by $w_t$ and its variance over $\epsilon$, given $x_0$ and $t$, by $w_t^2$. A constant multiple of the whole loss smaller than one has the same two effects. What a constant cannot change is how the gradient's second moment is spread across timesteps, and that is the quantity we study next.

For a minibatch of $B$ independently sampled examples $x_{i,0}$, timesteps $t_i$ and noises $\epsilon_i$, with $x_{i,t_i}=\sqrt{\bar\alpha_{t_i}}\,x_{i,0}+\sigma_{t_i}\epsilon_i$, the stochastic gradient of the weighted objective is
\begin{equation}
\hat g(\theta)
=
\frac{1}{B}
\sum_{i=1}^{B}
w(t_i,\lambda)
\nabla_\theta
\bigl\|
\epsilon_i-
\epsilon_\theta(x_{i,t_i},t_i)
\bigr\|^2.
\label{eq:minibatch}
\end{equation}
The weights change the relative contributions of different noise levels to this gradient, and so its direction as well as its length. To measure the change, fix $\theta$ and let $m_t=\mathbb E_{x_0,\epsilon}\bigl[\|\nabla_\theta\ell_t(\theta)\|^2\bigr]$ be the unweighted second moment at timestep $t$. Under uniform timestep sampling, timestep $t$ contributes in proportion to $w_t^2m_t$ to the second moment of the single-sample gradient, and normalizing gives
\begin{equation}
p_w(t)
=
\frac{w_t^2m_t}
{\sum_{s=1}^{T}w_s^2m_s},
\qquad t=1,\ldots,T.
\label{eq:gradient_distribution}
\end{equation}
This is a distribution over timesteps, and it does not change when all weights are multiplied by the same positive constant.

\begin{proposition}[Redistribution of gradient second moments]
\label{prop:share}
Suppose the timesteps are ordered from highest
to lowest SNR, and let $m_t\ge0$ satisfy
$\sum_{t=1}^{T}m_t>0$. For any positive
weight sequence $w=(w_1,\ldots,w_T)$, define
$p_w(t)$ by Eq.~\ref{eq:gradient_distribution}
and let $S_w(k)=\sum_{t=1}^{k}p_w(t)$, $k=1,\ldots,T$; write $S_{\mathrm{DDPM}}$ for $S_w$ with $w_t\equiv1$. Then $S_{cw}(k)=S_w(k)$ for every $c>0$.
Moreover, if $w'$ is another positive weight sequence and $w_t/w'_t$ is nondecreasing in $t$, then $S_w(k)\le S_{w'}(k)$ for all $k.$ In particular, for the smoothed-KL weights
and $\lambda_1\ge\lambda_2\ge0$,
\begin{equation}
S_{w(\cdot,\lambda_1)}(k)
\le
S_{w(\cdot,\lambda_2)}(k)
\le
S_{\mathrm{DDPM}}(k),
\qquad k=1,\ldots,T.
\label{eq:gradient_share_ordering}
\end{equation}
\end{proposition}

In the language of stochastic orders, Eq.~\ref{eq:gradient_share_ordering} says that $p_{w(\cdot,\lambda_1)}$ first-order stochastically dominates $p_{w(\cdot,\lambda_2)}$: a larger $\lambda$ moves the second moment toward later, lower-SNR timesteps. Put simply, the nearly clean timesteps become a smaller part of what the optimizer sees. Because $m_t$ is arbitrary, the direction of the shift holds for any architecture; its size depends on the $m_t$. Adaptive optimizers such as Adam, which scale their updates by coordinatewise squared gradients, can respond to such a shift. Appendix~\ref{rem:timestep_lr} works out what this implies for a minibatch update, where the mean gradients of different timesteps also enter.

In the trained models (Appendix~\ref{app:diagnostic}), the weighted methods have smaller gradient second moments than DDPM at high-SNR timesteps, and aggregate gradient-norm variability is similar across methods. Each model is measured at its own parameters, so these numbers complement Proposition~\ref{prop:share}, which holds at fixed parameters.

\section{Experiments}
\label{sec:experiments}

After the CIFAR-10 sweeps (Section~\ref{sec:lambda_ablation}), we test whether the density of noise levels explains the poor default threshold (Section~\ref{sec:density}), compare our weight with Min-SNR (Section~\ref{sec:matching}) and with uniform weighting in training speed (Section~\ref{sec:speed}), and check whether the findings carry over to the cosine schedule and to CelebA-64 (Sections~\ref{sec:cosine} and~\ref{sec:celeba64}).

All comparisons use a fixed compute budget: the same architecture, optimizer, schedule, exponential moving average (EMA), data order and iteration count for every method, so only the weighting changes. On CIFAR-10~\citep{krizhevsky2009learning} at $32\times32$ we use a U-Net with 128 base channels, multipliers $(1,2,2,2)$, two residual blocks per scale and self-attention at $16\times16$ ($\sim$33M parameters); AdamW with learning rate $2\times10^{-4}$, zero weight decay, cosine decay to $0$ and gradient-norm clipping at $1.0$; dropout $0.1$, random horizontal flips and bfloat16 mixed precision; the linear schedule ($\beta_1=10^{-4}$, $\beta_T=0.02$, $T=1000$); EMA decay $0.9999$; batch size 128; and 500 epochs ($\sim$195K iterations) at seeds 42, 123 and 456. FID~\citep{heusel2017gans} uses 50K samples from DDIM with 50 steps~\citep{song2021ddim} and is computed with torch-fidelity~\citep{obukhov2020torchfidelity} against the CIFAR-10 training set; DDPM 1000-step results are in Appendix~\ref{app:ddpm_fid}. All methods run at 122 ms per iteration.

Baselines are DDPM, Min-SNR with $\gamma\in\{2.5,5,10,20,50,100,200\}$ ($\gamma=2.5$ added later to test the matching rule), P2 at its default $k=1$, $\gamma_{\mathrm{P2}}=1$, and VDM~\citep{kingma2021variational} with the discrete approximation $(\SNR(t-1)-\SNR(t))/\SNR(t)$, with the weight at $t=1$ set to $0$. The P2 and VDM weights are normalized to unit mean over timesteps; Min-SNR and our weight are used as written. For our weight we swept $\lambda\in\{0.01,0.05,0.1,0.2,0.5,1.0\}$. Every configuration has three seeds, and we report mean $\pm$ standard deviation. Two-sample comparisons give the difference in means with its $95\%$ Welch interval and $p$-value, using Welch's approximate degrees of freedom (between 2 and 4 with three seeds per arm). We read comparisons through their intervals.

\subsection{Generation quality and sweeps}
\label{sec:lambda_ablation}

\begin{table}[t]
\centering
\small
\caption{CIFAR-10, linear schedule, FID (50K samples, DDIM-50), 3 seeds, mean $\pm$ std. Training timesteps are uniform except in the rows marked $\dagger$, which draw them from k-diffusion's density (Section~\ref{sec:density}). Gradient-norm statistics are in Appendix~\ref{app:diagnostic}.}
\label{tab:main_results}
\begin{tabular}{lc@{\hspace{3em}}lc@{\hspace{3em}}lc}
\toprule
Method & FID$\downarrow$ & Method & FID$\downarrow$ & Method & FID$\downarrow$ \\
\midrule
DDPM & $10.62\pm0.25$ & Min-SNR ($\gamma\!=\!2.5$) & $12.30\pm0.14$ & Ours ($\lambda\!=\!0.01$) & $10.05\pm0.14$ \\
P2 ($k\!=\!1,\gamma_{\mathrm{P2}}\!=\!1$) & $13.44\pm0.29$ & Min-SNR ($\gamma\!=\!5$) & $11.56\pm0.30$ & Ours ($\lambda\!=\!0.05$) & $11.27\pm0.28$ \\
VDM & $24.18\pm0.23$ & Min-SNR ($\gamma\!=\!10$) & $11.28\pm0.10$ & Ours ($\lambda\!=\!0.1$) & $11.57\pm0.53$ \\
 & & Min-SNR ($\gamma\!=\!20$) & $10.43\pm0.16$ & Ours ($\lambda\!=\!0.2$) & $11.94\pm0.16$ \\
 & & Min-SNR ($\gamma\!=\!50$) & $10.32\pm0.24$ & Ours ($\lambda\!=\!0.5$) & $12.25\pm0.04$ \\
 & & Min-SNR ($\gamma\!=\!100$) & $10.26\pm0.27$ & Ours ($\lambda\!=\!1.0$) & $12.97\pm0.22$ \\
 & & Min-SNR ($\gamma\!=\!200$) & $10.69\pm0.23$ & & \\
\midrule
DDPM$^\dagger$ & $8.97\pm0.43$ & & & Ours ($\lambda\!=\!0.2$)$^\dagger$ & $10.62\pm0.19$ \\
\bottomrule
\end{tabular}
\end{table}

Table~\ref{tab:main_results} collects the CIFAR-10 results. Mean FID improves monotonically as $\lambda$ decreases across our grid, and the best tested positive value, $\lambda=0.01$, lies at the boundary of the grid. DDPM ($\lambda=0$, $10.62\pm0.25$) is worse, but we tested no value between $0$ and $0.01$, so the optimum over $\lambda$ has not been located. The Min-SNR curve improves steeply from $\gamma=2.5$ to $\gamma=20$, changes little across $\gamma\in\{20,50,100\}$ ($10.43$ to $10.26$, a spread comparable to the seed standard deviations), and gets worse at $\gamma=200$. The $\gamma=5$ default recommended by~\citet{hang2023minsnr} is worse here ($11.56$ against $\sim10.3$ on the plateau).

Our best setting is $0.57$ FID below DDPM ($95\%$ interval $[-1.08,-0.06]$, Welch $t=-3.45$ on $3.1$ df, $p=0.04$). Because $\lambda$ was selected on this FID, the comparison is biased in favor of our weight, and under DDPM-1000 sampling the difference is $+0.07$ ($[-0.46,0.60]$; Appendix~\ref{app:ddpm_fid}). We therefore treat our weight and DDPM as on par in final FID.

\subsection{Does the sampling density explain the default threshold?}
\label{sec:density}

The thresholds used in practice fare poorly here. The k-diffusion default $\gamma'=1/\sigma_{\mathrm{data}}^2=4$ (Section~\ref{sec:relations}) maps to $\lambda=1/3$ under VP schedules, which Table~\ref{tab:main_results} brackets at $1.3$ to $1.6$ FID above DDPM. One explanation is resolution: \citet{hoogeboom2025simpler} shift the equivalent sigmoid toward lower log-SNR as resolution grows. Another is the density of sampled noise levels. For CIFAR-10, k-diffusion draws log-SNR from the cosine-schedule density~\citep{nichol2021improved} centered at $\log(1/\sigma_{\mathrm{data}}^2)\approx1.4$, and on that density $\gamma'=4$ leaves $10.0\%$ of the effective weight above $\log\SNR=2$, about what DDPM has on our schedule; under uniform $t$ it leaves $1.8\%$ (Appendix~\ref{app:edm}). If the density were the explanation, damping should stop hurting under k-diffusion's density.

We tested this directly. With everything else unchanged, we drew each training timestep by sampling log-SNR from k-diffusion's density and taking the nearest timestep of the linear schedule, and trained DDPM and our weight at $\lambda=0.2$ ($\gamma'=6$, the measured end of the bracket) with three seeds each (rows marked $\dagger$ in Table~\ref{tab:main_results}). The new density lowers FID for both weights, by $1.65$ ($[0.77,2.52]$) for DDPM and by $1.32$ ($[0.91,1.72]$) for ours, and DDPM at $8.97$ is the best FID in this paper. The penalty for damping stays: our weight is $1.65$ FID worse than DDPM ($[0.75,2.55]$, $p=0.011$), against $1.32$ ($[0.81,1.83]$) under uniform $t$, a change of $+0.33$ ($[-0.50,1.16]$). The density therefore does not explain the poor showing of the default threshold.

What orders these four runs is the share of the effective weight above $\log\SNR=2$: $40\%$, $12\%$, $11\%$ and $2\%$ go with FIDs of $8.97$, $10.62$, $10.62$ and $11.94$. At this resolution more weight at high SNR helped under either density, which points to resolution or parameterization rather than density as the reason a threshold tuned elsewhere does poorly here. Changing the density also improved FID by more than any weight we tried improved on uniform weighting ($0.57$ at best).

\subsection{Which Min-SNR threshold matches in FID?}
\label{sec:matching}

Neither the high-SNR match $\gamma=1/\lambda$ of Section~\ref{sec:relations} nor the one-point match $\gamma\approx1/(2\lambda)$ of Eq.~\ref{eq:local_match} predicts FID, so we compare each $\lambda$ with Min-SNR at both thresholds (Table~\ref{tab:matching}). Only the pair $\lambda=0.01$, $\gamma=100$ was planned before the runs; the rest is exploratory, with unadjusted intervals.

\begin{table}[t]
\centering
\small
\caption{Candidate matches on CIFAR-10 (linear, DDIM-50, 3 seeds each), all taken from Table~\ref{tab:main_results}. $\Delta$ is Ours minus Min-SNR with its $95\%$ Welch interval. Two intervals exclude $0$: the high-SNR match at $\lambda=0.05$ and the one-point match at $\lambda=0.2$.}
\label{tab:matching}
\begin{tabular}{lccccc}
\toprule
 & & \multicolumn{2}{c}{Min-SNR at $\gamma=1/\lambda$} & \multicolumn{2}{c}{Min-SNR at $\gamma=1/(2\lambda)$} \\
\cmidrule(lr){3-4}\cmidrule(lr){5-6}
$\lambda$ & Ours & FID & $\Delta$ $[95\%]$ & FID & $\Delta$ $[95\%]$ \\
\midrule
$0.2$  & $11.94\pm0.16$ & $11.56\pm0.30$ ($\gamma\!=\!5$)   & $+0.38$ $[-0.24,1.00]$ & $12.30\pm0.14$ ($\gamma\!=\!2.5$) & $-0.36$ $[-0.70,-0.02]$ \\
$0.1$  & $11.57\pm0.53$ & $11.28\pm0.10$ ($\gamma\!=\!10$)  & $+0.29$ $[-0.97,1.55]$ & $11.56\pm0.30$ ($\gamma\!=\!5$)  & $+0.01$ $[-1.08,1.10]$ \\
$0.05$ & $11.27\pm0.28$ & $10.43\pm0.16$ ($\gamma\!=\!20$)  & $+0.84$ $[0.27,1.41]$ & $11.28\pm0.10$ ($\gamma\!=\!10$) & $-0.01$ $[-0.62,0.60]$ \\
$0.01$ & $10.05\pm0.14$ & $10.26\pm0.27$ ($\gamma\!=\!100$) & $-0.21$ $[-0.77,0.35]$ & $10.32\pm0.24$ ($\gamma\!=\!50$) & $-0.27$ $[-0.76,0.22]$ \\
\bottomrule
\end{tabular}
\end{table}

The pre-planned pair, $\lambda=0.01$ against $\gamma=100$, differs by $-0.21$ FID ($[-0.77,0.35]$; $+0.16$, $[-0.37,0.69]$, under DDPM-1000 sampling, Appendix~\ref{app:ddpm_fid}). It sits where Min-SNR changes little with $\gamma$, so it cannot separate thresholds between $20$ and $100$. Larger $\lambda$ can. At $\lambda=0.05$, Min-SNR at $\gamma=20$ is $0.84$ FID better than our weight ($[0.27,1.41]$, $p=0.018$), against $-0.01$ ($[-0.62,0.60]$) at $\gamma=10$. At $\lambda=0.1$ both intervals are more than two FID points wide. At $\lambda=0.2$ our weight is $0.38$ worse than Min-SNR at $\gamma=5$ ($[-0.24,1.00]$) and $0.36$ better than at $\gamma=2.5$ ($[-0.70,-0.02]$, $p=0.044$), the run added to test the one-point match out of sample.

Inverting the Min-SNR FID curve over $\gamma\in\{2.5,5,10,20\}$ turns these comparisons into a posterior for the matching threshold $\gamma^\star$ (Table~\ref{tab:inversion} in Appendix~\ref{app:inversion}). At $\lambda=0.05$, $\Pr(\gamma^\star<1/\lambda)=0.98$ and the median of $\gamma^\star\lambda$ is $0.49$, as the one-point match of Eq.~\ref{eq:local_match} suggests. At $\lambda=0.2$ the out-of-sample test does not confirm it: the median of $\gamma^\star\lambda$ is $0.70$ and $\Pr(\gamma^\star<1/(2\lambda))=0.06$. At $\lambda=0.01$ the Min-SNR curve is too flat to separate the maps, and at $\lambda=0.1$ the seed spread is too large. Where the data are informative, the matching threshold lies between $1/(2\lambda)$ and $1/\lambda$, at about $0.5/\lambda$ for $\lambda=0.05$ and $0.7/\lambda$ for $\lambda=0.2$.

Published comparisons favor the soft clip: in HDiT, Soft-Min-SNR reaches FID $27.74$ against $35.62$ for Min-SNR at the same $\gamma=4$~\citep[Table~5]{crowson2024hourglass}, and~\citet[Table~3]{hang2025improved} find it ahead at two of three guidance scales. At $\gamma=1/\lambda$ our runs mostly go the other way: Min-SNR has the lower mean FID for $\lambda\ge0.05$, significantly at $0.05$ (Table~\ref{tab:matching}). Proposition~\ref{prop:asymptotic_minsnr} gives the sign. At a common threshold the soft weight damps more at every timestep, so it wins where more damping helps, as at higher resolutions~\citep{hoogeboom2025simpler}, and loses where less helps, as on CIFAR-10 (Table~\ref{tab:main_results}). It does not give the size of the HDiT gap, which its authors credit mainly to smoothness.

\subsection{Convergence speed}
\label{sec:speed}

\begin{figure}[t]
\centering
\includegraphics[width=0.9\linewidth]{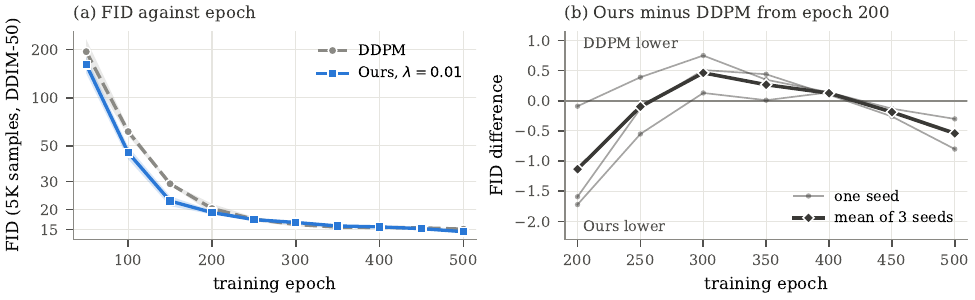}
\caption{CIFAR-10, linear schedule: 5K-sample DDIM-50 FID every 50 epochs for DDPM and Ours ($\lambda=0.01$), 3 seeds. (a) Mean and $\pm1$ std band. (b) Ours minus DDPM for each seed. By 3-seed medians with linear interpolation, Ours reaches FID 25, 23, 21 and 19 at epochs 143, 148, 172 and 207, against 174, 187, 200 and 231 for DDPM, and FID 17 and 16 at epochs 265 and 336, against 272 and 308.}
\label{fig:trajectory}
\end{figure}

Min-SNR was introduced to speed up training~\citep{hang2023minsnr}, so final FID is only part of the comparison. On CIFAR-10 with the linear schedule our weight leads DDPM early (Figure~\ref{fig:trajectory}), reaching every target from FID 25 to FID 19 earlier in each of the three seeds, with the largest margin at FID 23 (21\% fewer epochs by the 3-seed median). The lead does not last. At epochs 300, 350 and 400 DDPM has the lower FID in all three seeds, by $0.01$ to $0.75$, and it reaches FID 16 earlier; at epochs 450 and 500 ours is lower again in all three seeds, by $0.13$ to $0.80$. Which method is ahead therefore depends on when training stops. \citet[Table~1]{hang2023minsnr} report the same pattern for Min-SNR with $\epsilon$-prediction, whose FID lead over uniform weighting shrinks from about $1.2$ at 200K iterations to $0.07$ at 1M. On CelebA-64, DDPM has the lowest FID from epoch 100 to 250 in the one run per method we tracked (Section~\ref{sec:celeba64}).

\subsection{Cosine schedule}
\label{sec:cosine}

Under the cosine schedule of~\citet{nichol2021improved} (otherwise unchanged, 3 seeds), DDPM, Min-SNR-$\gamma\!=\!100$ and Ours-$\lambda\!=\!0.01$ reach FID $9.71\pm0.15$, $10.66\pm1.15$ and $9.90\pm0.84$: ours is $0.76$ below Min-SNR ($[-3.13,1.61]$) and $0.19$ above DDPM ($[-1.81,2.19]$), intervals several FID points wide. DDPM gains the most from this schedule, and both reweightings vary more across seeds under it (Ours $0.84$ vs $0.14$ on linear, Min-SNR $1.15$ vs $0.27$; variance-ratio $p\approx0.05$ and $0.10$).

\subsection{CelebA-64}
\label{sec:celeba64}

Finally, we test whether $\lambda=0.01$ and $\gamma=100$ transfer to a second dataset without re-tuning. CelebA-64 is the full aligned CelebA~\citep{liu2015celeba} (202{,}599 images), center-cropped to $178\times178$ and resized bicubically to $64\times64$. We train the same three methods on it under the cosine schedule, with attention at $\{16,8\}$ ($\sim$34M parameters), for 316 epochs ($\approx$500K iterations, batch 128) and 3 seeds, and compute FID-50K against the same prepared images.

Mean FIDs are $6.89\pm0.37$ for DDPM, $6.60\pm0.22$ for Min-SNR and $7.89\pm1.18$ for ours (Table~\ref{tab:celeba} in Appendix~\ref{app:celeba_curve}). Our weight is $1.28$ FID above Min-SNR on average (Welch $t=1.85$ on $2.1$ df, $p=0.20$, $95\%$ interval $[-1.52,\,4.09]$) and $1.00$ above DDPM ($p=0.28$, $[-1.64,\,3.64]$), much of it from one run (9.23), so the transfer of $\lambda=0.01$ to CelebA-64 is inconclusive. Along one run per method (seed 42, Figure~\ref{fig:c6_curve}), the two reweightings track each other more closely than either tracks DDPM, which has the lowest FID of the three from epoch 100 to 250. Each run took $43.1$ to $43.7$ hours on a single GPU.

\section{Discussion}
\label{sec:discussion}

The main message of the paper is that Soft-Min-SNR is what one obtains by comparing a predicted and a true noisy replica after Gaussian smoothing, and its threshold is the inverse of the smoothing variance. For a practitioner, the maps of Section~\ref{sec:relations} reduce the choice among Soft-Min-SNR, P2 with unit exponent and the sigmoid weighting to one number, with two cautions. Min-SNR matches our weight only at the highest SNRs; the threshold that matched FID was about $0.5/\lambda$ at $\lambda=0.05$ and $0.7/\lambda$ at $\lambda=0.2$, so we would start tuning between $1/(2\lambda)$ and $1/\lambda$. And the density of training noise levels deserves as much attention as the weight: on CIFAR-10, changing it improved FID by more than any weight improved on uniform weighting, and it left the ordering of the two weights we compared unchanged.

Against uniform weighting the results depend on the setting and the stage of training, consistent with Proposition~\ref{prop:share}: how far the weight moves the gradient's second moment depends on per-timestep moments $m_t$ that differ across settings and change during training.

Two recent accounts point the other way at high SNR. \citet{zhang2025heuristic} obtain the usual heuristic weighting of denoising score matching as a first-order approximation to an optimal weighting set by the heteroskedasticity of the objective, and \citet{gagneux2026training} justify weights that grow as the noise vanishes; on CIFAR-10, weighting uniform on the $\epsilon$-loss did best for them. For Gaussian data in VE, inverse-variance weighting on the $x_0$-loss is $1/\sigma^2+1/s^2$, close to the larger of $\SNR$ and $1/s^2$, while ours is $1/(\sigma^2+s^2)$, close to the smaller. The two answer different questions: inverse-variance weighting makes estimation efficient when the network can represent the target, and then every positive weight has the same minimizer; a divergence decides which errors to accept when it cannot. Our CIFAR-10 results, under both densities, also favor less damping.

\subsection{Limitations}
\label{sec:limitations}

The experiments use CIFAR-10 at $32\times32$ and CelebA-64 with U-Nets of about 33M and 34M parameters, and CIFAR-10 FID is around 10, above what EDM-class pipelines reach~\citep{karras2022elucidating}. Three seeds per configuration limit any single comparison, and on the cosine schedule, CelebA-64 and the k-diffusion density we tested only two or three settings. Testing the parameter maps on ImageNet, latent diffusion and EDM-style training (Appendix~\ref{app:edm}), and whether damping helps at higher resolution, are the natural next steps. On the theory side, the comparison is made one timestep at a time and is not a likelihood bound, and how the parameters reached by a network of limited capacity depend on $\lambda$ is not characterized.

\section*{Broader Impact Statement}

This work studies the training loss of diffusion models, which can also produce misleading synthetic media. It explains and connects existing reweightings, adds no new capability and releases no pretrained model. Code for all experiments is available from the corresponding author.

\clearpage
\appendix

\section{Proofs and Derivations}
\label{sec:proofs}

\subsection{Propositions~\ref{prop:contraction} and~\ref{prop:recovery}}

For Proposition~\ref{prop:contraction}, $K_\Lambda$ is a Markov kernel, so the data-processing inequality for KL gives $\KL(K_\Lambda q\,\|\,K_\Lambda p)\le\KL(q\,\|\,p)$. Equality at $\Lambda=0$ holds because $K_0$ is the identity.

For Proposition~\ref{prop:recovery}, write $D(\Lambda)=\KL(K_\Lambda q\,\|\,K_\Lambda p)$. If $\Lambda_1\preceq\Lambda_2$, then $\mathcal N(0,\Lambda_2)=\mathcal N(0,\Lambda_1)*\mathcal N(0,\Lambda_2-\Lambda_1)$, so $K_{\Lambda_2}=K_{\Lambda_2-\Lambda_1}K_{\Lambda_1}$, and applying the data-processing inequality to the Markov kernel $K_{\Lambda_2-\Lambda_1}$ gives $D(\Lambda_2)\le D(\Lambda_1)$. For the limit, let $\Lambda_n\to0$. If $X\sim q$ and $Z\sim\mathcal N(0,I)$ are independent, $K_{\Lambda_n}q$ is the law of $X+\Lambda_n^{1/2}Z$, which converges to $X$ in probability, so $K_{\Lambda_n}q\Rightarrow q$ weakly, and likewise $K_{\Lambda_n}p\Rightarrow p$. Relative entropy is jointly lower semicontinuous under weak convergence, which follows from the Donsker--Varadhan representation~\citep{donsker1983asymptotic} $\KL(q\|p)=\sup_f\{\mathbb{E}_q f-\log\mathbb{E}_p e^f\}$ over bounded continuous $f$, since each functional inside the supremum is continuous under weak convergence. Hence $\liminf_n D(\Lambda_n)\ge\KL(q\|p)$. Combined with $D(\Lambda_n)\le\KL(q\|p)$ from Proposition~\ref{prop:contraction}, $D(\Lambda_n)\to\KL(q\|p)$, including the case $\KL(q\|p)=+\infty$.

\subsection{Closed-form weight (Eq.~\ref{eq:loss_marginal})}

Both replica laws have covariance $\sigma_t^2I$ and means differing by $\delta=\sqrt{\bar\alpha_t}(x_0-\hat x_0)$. Since $x_0=(x_t-\sigma_t\epsilon)/\sqrt{\bar\alpha_t}$ and $\hat x_0=(x_t-\sigma_t\epsilon_\theta)/\sqrt{\bar\alpha_t}$, $\delta=\sigma_t(\epsilon_\theta-\epsilon)$, so $\KL(q_t\|p_{\theta,t})=\|\delta\|^2/(2\sigma_t^2)=\frac12\|\epsilon-\epsilon_\theta\|^2$, which is $L_{\mathrm{simple}}$ up to the factor $\frac12$. With $\Lambda=\lambda I$ both covariances become $(\sigma_t^2+\lambda)I$, and
\[
\sKL(q_t\|p_{\theta,t})=\frac{\|\delta\|^2}{2(\sigma_t^2+\lambda)}=\frac{\sigma_t^2}{\sigma_t^2+\lambda}\cdot\frac{\|\delta\|^2}{2\sigma_t^2}=w(t,\lambda)\,\KL(q_t\|p_{\theta,t}).
\]

\subsection{Proposition~\ref{prop:softminsnr}}

For VP schedules $\sigma_t^2=1-\bar\alpha_t$ and $\SNR(t)=(1-\sigma_t^2)/\sigma_t^2$. With $\gamma'=(1+\lambda)/\lambda$, multiplying numerator and denominator of $w_{\mathrm{SMS}}$ by $\lambda\sigma_t^2$ gives
\[
w_{\mathrm{SMS}}(t,\gamma')=\frac{(1+\lambda)\sigma_t^2}{(1+\lambda)\sigma_t^2+\lambda(1-\sigma_t^2)}=\frac{(1+\lambda)\sigma_t^2}{\sigma_t^2+\lambda}=(1+\lambda)\,w(t,\lambda).
\]
The factor $1/(1+\lambda)$ does not depend on $t$, so gradient descent on the two losses differs only by a learning-rate rescaling. In the VE parameterization $x_\sigma=x_0+\sigma\epsilon$, take $q_\sigma=\mathcal N(x_0,\sigma^2I)$ and $p_{\theta,\sigma}=\mathcal N(\hat x_0,\sigma^2I)$ with $\hat x_0=x_\sigma-\sigma\epsilon_\theta$. Then $\delta=\sigma(\epsilon_\theta-\epsilon)$, Eq.~\ref{eq:closure} gives $\sKL(q_\sigma\|p_{\theta,\sigma})=\|\delta\|^2/(2(\sigma^2+\lambda))=\frac{\sigma^2}{\sigma^2+\lambda}\cdot\frac12\|\epsilon-\epsilon_\theta\|^2$, and with $\SNR=1/\sigma^2$ this weight is $1/(1+\lambda\,\SNR)=w_{\mathrm{SMS}}(\sigma,1/\lambda)$.

\subsection{Proposition~\ref{prop:asymptotic_minsnr}}
\label{app:minsnr_proof}

From Eq.~\ref{eq:snr_form}, $w(t,\lambda)=1/(1+x/\bar\alpha_t)$ with $x=\lambda\,\SNR(t)>0$. Since $0<\bar\alpha_t\le1$, $x/\bar\alpha_t\ge x$ and so $w(t,\lambda)\le1/(1+x)$. For $x>0$, $1/(1+x)<1$ and $1/(1+x)<1/x$, hence $1/(1+x)<\min(1,1/x)$. With $\gamma=1/\lambda$, $\min(1,1/x)=\min(1,\gamma/\SNR(t))=\min(\SNR(t),\gamma)/\SNR(t)$, the Min-SNR weight. For the ratio, where $x\ge1$ the Min-SNR weight is $\gamma/\SNR(t)=1/x$, so $w(t,\lambda)\big/(\gamma/\SNR(t))=x/(1+x/\bar\alpha_t)=\bar\alpha_tx/(\bar\alpha_t+x)<\bar\alpha_t$; on VP and VE schedules $\bar\alpha_t\to1$ as $\SNR(t)\to\infty$, so the ratio tends to $1$. At $\SNR(t)=\gamma$ we have $x=1$ and $w(t,\lambda)=1/(1+1/\bar\alpha_t)\le1/2$. For the relative error under a VP schedule, write $s=\SNR(t)$. Eq.~\ref{eq:p2} gives $w(t,\lambda)=1/(1+\lambda+\lambda s)$, and where $s>\gamma$ the Min-SNR weight is $\gamma/s=1/(\lambda s)$, so $1-w(t,\lambda)\,s/\gamma=(1+\lambda)/(1+\lambda+\lambda s)$. Since $s=(1-\sigma_t^2)/\sigma_t^2$, $1+\lambda+\lambda s=(\sigma_t^2+\lambda)/\sigma_t^2$, and the relative error is $(1+\lambda)\sigma_t^2/(\sigma_t^2+\lambda)$, which equals the expression in the proposition, is at most $(1+\lambda)\sigma_t^2/\lambda$, and tends to $0$ as $\sigma_t^2\to0$.

The average-weight comparison quoted after Eq.~\ref{eq:local_match} solves $\frac1T\sum_t\min(\SNR(t),\gamma)/\SNR(t)=\frac1T\sum_t w(t,\lambda)$ for $\gamma$ on the linear schedule of Section~\ref{sec:experiments}; for $\lambda\in\{0.01,0.05,0.1,0.2\}$ the solutions are $\gamma\approx14.2,\,2.1,\,0.84,\,0.29$, i.e.\ $\gamma\lambda\in[0.06,0.14]$.

\subsection{Local curvature (Section~\ref{sec:theory})}

As a function of the predicted mean, the auxiliary KL is $f(\mu_\theta)=\frac12\delta^\top\Sigma_{\mathrm{eff}}^{-1}\delta$ with $\delta=\sqrt{\bar\alpha_t}\,x_0-\mu_\theta$ and $\Sigma_{\mathrm{eff}}=\Sigma_t+\Lambda$, and it has Hessian $\Sigma_{\mathrm{eff}}^{-1}$, so the gradient Lipschitz constant is $\|\Sigma_{\mathrm{eff}}^{-1}\|_2$. For $\Lambda=\lambda I$, $\Sigma_{\mathrm{eff}}=(\sigma_t^2+\lambda)I$ and $L_\Lambda=1/(\sigma_t^2+\lambda)\le1/\sigma_t^2=L$.

\subsection{Proposition~\ref{prop:share}}

Scale invariance is immediate, since $c^2$ cancels between numerator and denominator. For the ordering, let $a_t=(w_t/w'_t)^2$, which is nonnegative and nondecreasing in $t$, and $b_t=w'^2_t m_t\ge0$. Fix $k<T$ (for $k=T$ both shares equal $1$) and write $B_1=\sum_{t\le k}b_t$, $B_2=\sum_{t>k}b_t$, $A_1=\sum_{t\le k}a_tb_t$ and $A_2=\sum_{t>k}a_tb_t$, so that $S_w(k)=A_1/(A_1+A_2)$ and $S_{w'}(k)=B_1/(B_1+B_2)$. Monotonicity of $a_t$ gives $A_1\le a_kB_1$ and $A_2\ge a_{k+1}B_2\ge a_kB_2$. Since $x\mapsto x/(x+y)$ is increasing in $x$ and decreasing in $y$ for $x,y\ge0$ with $x+y>0$,
\[
S_w(k)=\frac{A_1}{A_1+A_2}\le\frac{a_kB_1}{a_kB_1+a_kB_2}=S_{w'}(k)
\]
whenever $a_k>0$, which holds because $w$ and $w'$ are positive. For the particular case, $w(t,\lambda_1)/w(t,\lambda_2)=(\sigma_t^2+\lambda_2)/(\sigma_t^2+\lambda_1)$ has derivative $(\lambda_1-\lambda_2)/(\sigma_t^2+\lambda_1)^2\ge0$ in $\sigma_t^2$, and $\sigma_t^2$ increases with $t$. Taking $\lambda_2=0$ gives $w(t,0)=1$, the DDPM weight.

\subsection{Per-timestep learning rate under SGD and Adam}
\label{rem:timestep_lr}

Under plain SGD, applying $w(t,\lambda)$ is equivalent to training the unweighted loss with a timestep-dependent learning rate $\eta_t=w(t,\lambda)\eta$. Under Adam this equivalence breaks. The first- and second-moment estimates are formed from the minibatch gradient in Eq.~\ref{eq:minibatch}, and a global factor $c$ cancels between them (up to $\varepsilon$). A weight that differs across the timesteps in a batch, however, changes the relative contribution of different timesteps to both moments. SNR-based reweighting therefore keeps an effect under optimizers that already adapt step sizes per coordinate, whereas a constant multiplier does not.

What Proposition~\ref{prop:share} says about Adam is narrower. Fix $\theta$ and a coordinate $j$, let $G_{ij}=w(t_i,\lambda)\,\partial_j\ell_{t_i}(\theta)$, $i=1,\dots,B$, be coordinate $j$ of independent single-sample weighted gradients with $t_i$ uniform, and let $\hat g_j=B^{-1}\sum_{i=1}^BG_{ij}$. Then $\mathbb E[\hat g_j^2]=B^{-1}\,\mathbb E[G_{1j}^2]+(B-1)B^{-1}(\mathbb E[G_{1j}])^2$, since $\mathbb E[\hat g_j^2]=\mathrm{Var}(\hat g_j)+(\mathbb E\hat g_j)^2=B^{-1}(\mathbb E[G_{1j}^2]-(\mathbb E G_{1j})^2)+(\mathbb E G_{1j})^2$. With $m_t$ the second moment of the unweighted coordinate gradient at step $t$, the first term is $B^{-1}T^{-1}\sum_tw_t^2m_t$, and Proposition~\ref{prop:share} says that larger $\lambda$ lowers the share of it that comes from any high-SNR block $\{1,\dots,k\}$. The second term is $(T^{-1}\sum_tw_t\mu_{t,j})^2$, where $\mu_{t,j}$ is the mean of the unweighted coordinate gradient at step $t$. It contains products $\mu_{t,j}\mu_{t',j}$ between timesteps, which can have either sign, so it has no timestep decomposition of the same kind, and averaging over more batches estimates it rather than removing it. Adam's second-moment estimate is moreover an exponential average of $\hat g_j^2$ along the optimization path, so $\theta$ changes between the averaged terms. A statement about the timestep composition of Adam's normalizer would need information about the mean gradients $\mu_{t,j}$ as well as the second moments $m_t$, and we do not have one.

\subsection{Proposition~\ref{prop:predictive}}
\label{app:predictive}

Write $q(\cdot\mid x_0)$, $q(\cdot\mid x_t)$ and $p(\cdot\mid x_t)$ for $q_{\lambda,t}(\cdot\mid x_0)$, $q_{\lambda,t}(\cdot\mid x_t)$ and $p_{\theta,\lambda,t}(\cdot\mid x_t)$. Each is a density with respect to Lebesgue measure, since each contains Gaussian noise of variance $\sigma_t^2+\lambda>0$. The replica noise $\eta$ and the measurement noise $\xi$ are independent of $(X_0,\epsilon)$, so $Y_t$ and $X_t$ are conditionally independent given $X_0$ and $q(y\mid x_0)=q(y\mid x_0,x_t)$. Hence
\[
\log\frac{q(Y_t\mid X_0)}{p(Y_t\mid X_t)}=\log\frac{q(Y_t\mid X_0,X_t)}{q(Y_t\mid X_t)}+\log\frac{q(Y_t\mid X_t)}{p(Y_t\mid X_t)}.
\]
Taking expectations over $(X_0,X_t,Y_t)$, the left side gives the left side of Eq.~\ref{eq:predictive}, the first term on the right gives $I(X_0;Y_t\mid X_t)$ and the second gives $\mathbb E\,\KL\bigl(q(\cdot\mid X_t)\,\|\,p(\cdot\mid X_t)\bigr)$. All three are finite. The left side equals $\bar\alpha_t\,\mathbb E\|X_0-\hat x_0(X_t,t)\|^2/(2(\sigma_t^2+\lambda))$, which is finite under the moment conditions and, since $\bar\alpha_t\|x_0-\hat x_0\|^2=\sigma_t^2\|\epsilon-\epsilon_\theta\|^2$, equals $\frac12\,\mathbb E\,L_t^\lambda$; by convexity of KL, $\KL\bigl(q(\cdot\mid x_t)\,\|\,p(\cdot\mid x_t)\bigr)\le\mathbb E\bigl[\KL\bigl(q(\cdot\mid X_0)\,\|\,p(\cdot\mid x_t)\bigr)\mid X_t=x_t\bigr]$, so the second term is bounded by the left side; and the mutual information is the nonnegative difference of the two.

The mutual information involves only the data and the noise, not $\theta$. For $\lambda_1\le\lambda_2$, the observation at $\lambda_2$ is the observation at $\lambda_1$ plus independent Gaussian noise of variance $\lambda_2-\lambda_1$, so $X_0\to Y_t^{(\lambda_1)}\to Y_t^{(\lambda_2)}$ is a Markov chain given $X_t$, and the conditional data-processing inequality gives $I(X_0;Y_t^{(\lambda_2)}\mid X_t)\le I(X_0;Y_t^{(\lambda_1)}\mid X_t)$.

For the minimizer, the first term on the right of Eq.~\ref{eq:predictive} does not depend on $\hat x_0$, and
\[
\mathbb E\,\KL\bigl(q(\cdot\mid X_t)\,\|\,p(\cdot\mid X_t)\bigr)=\frac{\mathbb E\|Y_t-\sqrt{\bar\alpha_t}\,\hat x_0(X_t,t)\|^2}{2(\sigma_t^2+\lambda)}+c,
\]
where $c$, which collects the conditional entropy of $Y_t$ given $X_t$ and the Gaussian normalizing constant, does not depend on $\hat x_0$. Over measurable functions of $X_t$ this is minimized by $\sqrt{\bar\alpha_t}\,\hat x_0=\mathbb E[Y_t\mid X_t]=\sqrt{\bar\alpha_t}\,\mathbb E[X_0\mid X_t]$, since $\sigma_t\eta+\sqrt\lambda\,\xi$ has mean zero and is independent of $X_t$. The relation $\hat x_0=(x_t-\sigma_t\epsilon_\theta)/\sqrt{\bar\alpha_t}$ turns this into $\epsilon_\theta(x_t,t)=\mathbb E[\epsilon\mid X_t=x_t]$. None of this depends on $\lambda$, and because the full loss sums the per-timestep losses with positive weights, the minimizer of $\mathcal L_\lambda$ over unrestricted measurable predictors is the same for every $\lambda$. This is a statement about the population target only. When $\epsilon_\theta$ ranges over a restricted class that does not contain the posterior mean at every timestep, the minimizer within the class trades off errors across timesteps according to $w(t,\lambda)$ and generally depends on $\lambda$. The parameters reached by finite stochastic optimization also depend on $\lambda$ through the optimization path.

\section{Smoothed Score and Tweedie's Formula}
\label{app:tweedie}

The smoothed KL compares two distributions after both are convolved with the same Gaussian. This appendix shows what the convolution does to a density's score, which explains the smoothed score fields in Figure~\ref{fig:toy_scores} and the denoising-score-matching reading in Section~\ref{sec:method}.

\begin{theorem}[Noise-averaged score]
\label{thm:score}
Let $\Lambda\succ0$, let $p$ be a continuously differentiable density with enough decay that boundary terms vanish under integration by parts, and let $X=Y+V$ with $Y\sim p$ and $V\sim\mathcal N(0,\Lambda)$ independent, so that $X\sim K_\Lambda p$. Then
\[
\nabla\log(K_\Lambda p)(x)=\mathbb{E}_{Y\sim\pi(\cdot\mid x)}\bigl[\Lambda^{-1}(Y-x)\bigr]=\mathbb{E}\bigl[\nabla\log p(Y)\mid X=x\bigr],
\]
where $\pi(y\mid x)\propto p(y)\,\mathcal N(y;x,\Lambda)$ is the conditional density of $Y$ given $X=x$.
\end{theorem}

\emph{Proof.} Write $(K_\Lambda p)(x)=\int p(y)\phi_\Lambda(y-x)\,dy$ with $\phi_\Lambda$ the $\mathcal N(0,\Lambda)$ density. Since $\nabla_x\phi_\Lambda(y-x)=\Lambda^{-1}(y-x)\phi_\Lambda(y-x)$,
\[
\nabla_x\log(K_\Lambda p)(x)=\frac{\int p(y)\,\Lambda^{-1}(y-x)\,\phi_\Lambda(y-x)\,dy}{(K_\Lambda p)(x)}=\mathbb{E}_{\pi(\cdot\mid x)}\bigl[\Lambda^{-1}(Y-x)\bigr].
\]
Integrating by parts in the numerator moves the derivative onto $p$ and gives $\mathbb{E}_{\pi(\cdot\mid x)}[\nabla\log p(Y)]$. $\square$

This is Tweedie's formula~\citep{robbins1956empirical,efron2011tweedie}. Applied to the replica law $q_t(\cdot\mid x_0)$, $\nabla\log(K_\Lambda q_t)$ is the score of a Gaussian with mean coefficient $\sqrt{\bar\alpha_t}$ and covariance $\sigma_t^2+\lambda$, which is not the law $q_s(\cdot\mid x_0)$ at any other timestep $s$ of the same schedule. The trained model still learns a standard noise predictor: by Proposition~\ref{prop:predictive} its unrestricted population target is $\mathbb E[\epsilon\mid X_t]$ for every $\lambda$, and the smoothed score is a way to read the reweighted loss, not a training target. Figure~\ref{fig:toy_scores} in Section~\ref{sec:method} illustrates the operator on a toy mixture; it does not show that training with $w(t,\lambda)$ learns a more heavily smoothed score. The theorem also does not make the smoothed score small: for $p=\mathcal N(0,vI)$ it gives $-x/(v+\lambda)$, unbounded in $\|x\|$.

\section{Per-timestep Gradient Diagnostic}
\label{app:diagnostic}

\begin{figure}[ht]
\centering
\includegraphics[width=0.62\linewidth]{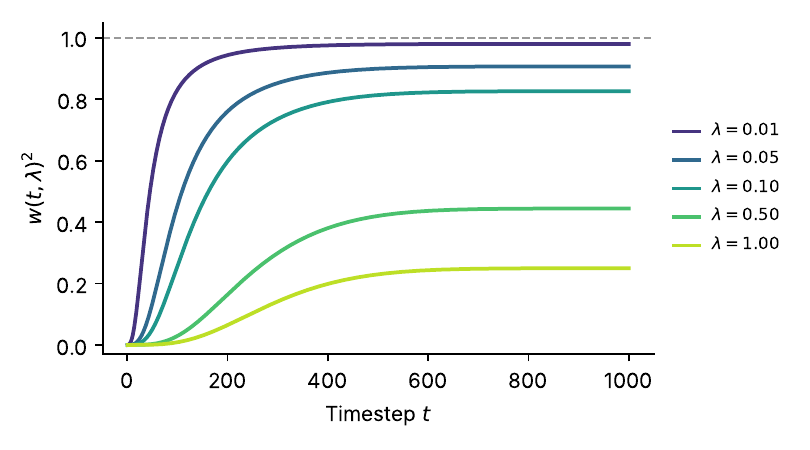}
\caption{Squared weight $w(t,\lambda)^2$, the variance factor in Section~\ref{sec:theory}, across timesteps for several $\lambda$; the dashed line at $1$ is DDPM. The weight itself, which equals the curvature ratio $L_\Lambda/L$ of Section~\ref{sec:theory}, is plotted in Figure~\ref{fig:weight_schedule} (left). Both are smallest at small $t$.}
\label{fig:theory_predictions}
\end{figure}

Figure~\ref{fig:theory_predictions} shows the analytic reference values. We measured the per-timestep ratio $\mathbb{E}\|g_t^{\mathrm{method}}\|^2/\mathbb{E}\|g_t^{\mathrm{DDPM}}\|^2$, where $g_t$ is the gradient of a method's weighted loss on a batch at timestep $t$, over 20 batches per timestep at each method's own trained parameters, at every tenth timestep below $200$. For Ours ($\lambda=0.1$) the empirical ratio is $0.213$ against a reference $w(t,\lambda)^2=0.253$; for Min-SNR ($\gamma=5$) it is $0.354$ against $0.455$. Each method is evaluated at its own trained $\theta$, so these are empirical comparisons between trained models: they neither verify the fixed-$\theta$ variance identity of Section~\ref{sec:theory} nor establish an ordering from it, because the unweighted per-timestep gradients differ between the parameters being compared. The value $w(t,\lambda)^2$ serves only as a reference. The measurement was made at $\lambda=0.1$ during the exploratory ablation. Since $w(t,\lambda)$ decreases as $\lambda$ grows, larger $\lambda$ damps more, and the reference damping at $\lambda=0.01$ is weaker than at $\lambda=0.1$; we do not report the corresponding measurement at $\lambda=0.01$. The coefficient of variation (CoV) of the gradient norm over the last 100 epochs lies between $0.048$ and $0.068$ (DDPM $0.050$; Min-SNR with $\gamma=5$, $0.049$; Ours with $\lambda=0.01$, $0.055$; P2 $0.068$; VDM $0.048$); VDM's training-loss CoV is higher ($0.045$ vs $0.005$ to $0.009$), consistent with the concentration of its weight on the smallest timesteps (at $t=2$ the normalized weight is $59$). The number of spikes (epochs where the gradient norm exceeds three times its 20-epoch running median) is $0$ for DDPM, VDM and Ours and $0.7$ on average for Min-SNR and P2. In these trained models, then, the weighted methods show smaller small-$t$ gradient second moments than DDPM but similar aggregate gradient-norm variability over training.

\section{Pseudocode and Weight Schedule}
\label{app:pseudocode}

\begin{figure}[ht]
\centering
\includegraphics[width=\linewidth]{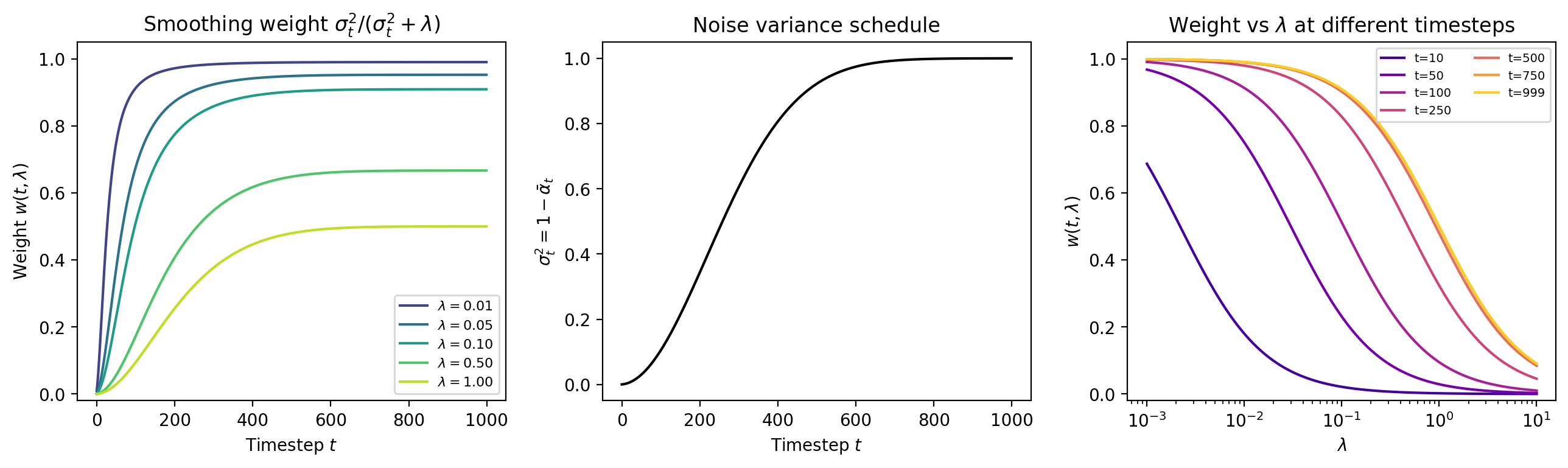}
\caption{Left: $w(t,\lambda)$ across timesteps for several $\lambda$. Center: the $\sigma_t^2$ schedule. Right: $w$ as a function of $\lambda$ at selected timesteps. The weight is smallest at small $t$, where the unweighted loss is steepest.}
\label{fig:weight_schedule}
\end{figure}

Algorithm~1 gives the training loop, and Figure~\ref{fig:weight_schedule} plots the weight schedule.

\begin{algorithm}[H]
\caption{DDPM training with the smoothed-KL weight.}
\begin{algorithmic}[1]
\State Input: data $p_0(x_0)$, schedule $\{\beta_t\}_{t=1}^T$, model $\epsilon_\theta$, smoothing $\lambda\ge0$.
\State Precompute $w(t)=\sigma_t^2/(\sigma_t^2+\lambda)$ for $t=1,\dots,T$.
\For{each training step}
  \State Sample $x_0\sim p_0$, $t\sim\mathrm{Uniform}\{1,\dots,T\}$, $\epsilon\sim\mathcal N(0,I)$.
  \State $x_t=\sqrt{\bar\alpha_t}\,x_0+\sqrt{1-\bar\alpha_t}\,\epsilon$.
  \State $\mathcal L=w(t)\cdot\|\epsilon-\epsilon_\theta(x_t,t)\|^2$.
  \State Update $\theta$ with the optimizer on $\mathcal L$.
\EndFor
\end{algorithmic}
\end{algorithm}

\section{Fourier Interpretation}
\label{app:fourier}

The Fourier transform of $K_\Lambda p$ is $\widehat{K_\Lambda p}(\omega)=\exp(-\frac12\omega^\top\Lambda\omega)\,\hat p(\omega)$. For $\Lambda=\lambda I$ this is a Gaussian low-pass filter with bandwidth of order $1/\sqrt\lambda$, so the smoothed KL mainly measures the low-frequency part of the density mismatch in $\mathbb{R}^d$. With isotropic $\Lambda$ the filter acts uniformly across image coordinates; it smooths the probability density, not individual images.

\section{CelebA-64 Results by Seed and FID Trajectory}
\label{app:celeba_curve}

\begin{figure}[ht]
\centering
\includegraphics[width=0.75\linewidth]{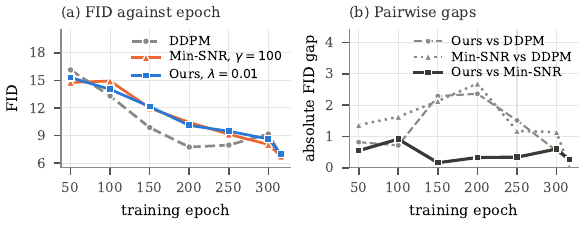}
\caption{CelebA-64, seed 42. (a) FID against epoch (10K samples; final point 50K). (b) Pairwise absolute FID gaps: $|$Ours$-$Min-SNR$|$ averages $0.45$ (max $0.91$), against $1.21$ and $1.45$ for the gaps to DDPM.}
\label{fig:c6_curve}
\end{figure}

Table~\ref{tab:celeba} gives the CelebA-64 FIDs by seed, and Figure~\ref{fig:c6_curve} tracks one run per method (seed 42), both summarized in Section~\ref{sec:celeba64}.

\begin{table}[ht]
\centering
\small
\caption{CelebA-64 FID-50K by seed, cosine schedule, 316 epochs. Min-SNR uses $\gamma=100$ and ours $\lambda=0.01$, both reused from CIFAR-10 without re-tuning.}
\label{tab:celeba}
\begin{tabular}{lcccc}
\toprule
Method & 42 & 123 & 456 & mean $\pm$ std \\
\midrule
DDPM & 6.78 & 6.58 & 7.30 & $6.89\pm0.37$ \\
Min-SNR & 6.75 & 6.35 & 6.71 & $6.60\pm0.22$ \\
Ours & 7.01 & 9.23 & 7.42 & $7.89\pm1.18$ \\
\bottomrule
\end{tabular}
\end{table}

\section{DDPM-1000 FID Scores}
\label{app:ddpm_fid}

The main tables use DDIM-50. Here we recompute FID with 1000-step DDPM sampling on 10K samples from the same trained checkpoints (Table~\ref{tab:ddpm1000}), including the three Min-SNR-$\gamma\!=\!100$ checkpoints of the pre-planned pair. Absolute FIDs are lower under this sampler, and the 10K sample size makes each FID noisier than the 50K DDIM-50 numbers. For the pre-planned pair, Ours minus Min-SNR is $+0.16$ ($7.87\pm0.26$ vs $7.71\pm0.15$, $95\%$ interval $[-0.37,0.69]$, Welch $p=0.42$), and Ours minus DDPM is $+0.07$ ($7.87\pm0.26$ vs $7.80\pm0.15$, $[-0.46,0.60]$, $p=0.71$), against $-0.57$ under DDIM-50. Neither interval excludes differences of about half an FID point. Min-SNR at the $\gamma=5$ default ($9.50\pm0.03$) has a higher FID than DDPM, Ours and Min-SNR at $\gamma=100$.

\begin{table}[ht]
\centering
\small
\caption{CIFAR-10 FID with DDPM-1000 sampling, 10K samples, 3 seeds (mean $\pm$ std).}
\label{tab:ddpm1000}
\begin{tabular}{lc}
\toprule
Method & FID$\downarrow$ \\
\midrule
DDPM & $7.80\pm0.15$ \\
Ours ($\lambda=0.01$) & $7.87\pm0.26$ \\
Min-SNR ($\gamma=100$, paired with $\lambda=0.01$) & $7.71\pm0.15$ \\
Min-SNR ($\gamma=5$, default) & $9.50\pm0.03$ \\
P2 ($k=1,\gamma_{\mathrm{P2}}=1$) & $12.41\pm0.20$ \\
VDM & $19.56\pm0.52$ \\
\bottomrule
\end{tabular}
\end{table}

\section{EDM and the Smoothed-KL Family}
\label{app:edm}

EDM~\citep{karras2022elucidating} trains with three coupled choices: a lognormal density over the noise level, $\log\sigma\sim\mathcal N(P_{\mathrm{mean}},P_{\mathrm{std}}^2)$; a per-sample multiplier on the denoiser loss; and a preconditioning that fixes how the network's output is parameterized at each noise level. In the variance-exploding parameterization $x_\sigma=x_0+\sigma\epsilon$ (so $\SNR=1/\sigma^2$) the multiplier on the denoiser loss is $(\sigma^2+\sigma_{\mathrm{data}}^2)/(\sigma\,\sigma_{\mathrm{data}})^2$, and since the denoiser output $D_\theta$ and the implied noise prediction $\hat\epsilon=(x_\sigma-D_\theta)/\sigma$ satisfy $\|D_\theta-x_0\|^2=\sigma^2\|\hat\epsilon-\epsilon\|^2$ the multiplier on the $\epsilon$-loss is $1+1/(\sigma_{\mathrm{data}}^2\SNR)$, which tends to $1$ at high SNR.

The sampling density and the multiplier are not separate effects on the objective. A loss estimated with noise levels drawn from one density equals, in expectation, the same loss estimated with noise levels drawn from any other density of full support and multiplied by the density ratio. \citet{kingma2023understanding} use this importance-weighting argument to write diffusion objectives as integrals over log-SNR with an effective weight (their Section~3.1), and their Table~1 includes EDM: with $\lambda_{\mathrm{KG}}=\log\SNR$, the default $P_{\mathrm{mean}}=-1.2$, $P_{\mathrm{std}}=1.2$ gives $\lambda_{\mathrm{KG}}\sim\mathcal N(2.4,2.4^2)$, and the effective weight per unit log-SNR on the $\epsilon$-loss is proportional to $\mathcal N(\lambda_{\mathrm{KG}};2.4,2.4^2)\,(e^{-\lambda_{\mathrm{KG}}}+\sigma_{\mathrm{data}}^2)$. Because $\sigma$ is a function of SNR, this is a function of SNR. On any given schedule the same effective weight is also produced by uniform-$t$ sampling with the per-sample multiplier equal to the effective weight divided by the density of $\log\SNR(t)$ under uniform $t$, which is again a function of $\SNR(t)$. Figure~\ref{fig:weights}b plots EDM's effective weight; it is unimodal, peaking near $\log\SNR\approx-3.3$, and so not monotone in log-SNR.

The bare per-sample multiplier, the sampling density over noise levels and their product, the effective weight per unit log-SNR, are therefore different objects. Table~\ref{tab:weights} lists per-sample multipliers, the propositions of Section~\ref{sec:relations} compare such multipliers at a common, uniform-$t$ density, and Figure~\ref{fig:weights}b compares effective weights. EDM's weighting is representable in the same form, so it can be compared with ours by converting both to effective weights per unit log-SNR. What does not carry over automatically is the comparison of complete training procedures. Two procedures with the same effective weight have the same expected loss as a function of the denoiser, but their stochastic gradients have different variances, and EDM's preconditioning changes how the denoiser is parameterized at each noise level. The reciprocal maps of Section~\ref{sec:relations} were derived for $\epsilon$-prediction networks trained with uniform-$t$ sampling, and they do not automatically extend to the complete EDM procedure. Comparing with EDM empirically would require matching the sampling density, the multiplier and the parameterization together.

The same distinction explains why a threshold tuned with one density need not transfer to another. Table~\ref{tab:density_share} gives the share of the effective weight per unit log-SNR that lies above $\log\SNR=2$, the high-SNR end of the range, for several combinations of density and multiplier. Under the cosine density that k-diffusion uses for CIFAR-10, $40\%$ of sampled noise levels lie above $\log\SNR=2$, against $11\%$ under uniform $t$ on the linear schedule. Across the configurations we trained, FID falls as this share rises, with one exception: the light damping at $\lambda=0.01$ is slightly better than DDPM despite a smaller share. Section~\ref{sec:density} tests whether the density explains the poor showing of the k-diffusion default and finds that it does not. With the cut at $\log\SNR=1$ the ordering of the rows with FIDs is the same; at $3$ the two rows with FID $10.62$ swap places.

\begin{table}[ht]
\centering
\small
\caption{Share of the effective weight per unit log-SNR above $\log\SNR=2$, on the $\epsilon$-loss, with SNR computed for data scaled to $[-1,1]$. ``Cosine'' is the log-SNR density of the cosine schedule centered at $\log(1/\sigma_{\mathrm{data}}^2)$ that k-diffusion uses for CIFAR-10, ``lognormal'' is EDM's default, and ``uniform $t$'' is on the linear schedule. FIDs are from Table~\ref{tab:main_results}.}
\label{tab:density_share}
\begin{tabular}{llcc}
\toprule
Density & Multiplier & Share above $\log\SNR=2$ & FID \\
\midrule
Cosine (k-diffusion) & none (DDPM) & $40.3\%$ & $8.97$ \\
Cosine (k-diffusion) & ours, $\lambda=0.2$ & $12.3\%$ & $10.62$ \\
Cosine (k-diffusion) & Soft-Min-SNR, $\gamma'=4$ & $10.0\%$ & \\
Lognormal (EDM) & EDM & $8.7\%$ & \\
Uniform $t$ & none (DDPM) & $10.8\%$ & $10.62$ \\
Uniform $t$ & ours, $\lambda=0.01$ & $7.5\%$ & $10.05$ \\
Uniform $t$ & ours, $\lambda=0.05$ & $4.7\%$ & $11.27$ \\
Uniform $t$ & ours, $\lambda=0.1$ & $3.4\%$ & $11.57$ \\
Uniform $t$ & ours, $\lambda=0.2$ & $2.4\%$ & $11.94$ \\
Uniform $t$ & ours, $\lambda=1/3$ ($\gamma'=4$) & $1.8\%$ & \\
Uniform $t$ & ours, $\lambda=0.5$ & $1.5\%$ & $12.25$ \\
Uniform $t$ & ours, $\lambda=1$ & $1.1\%$ & $12.97$ \\
\bottomrule
\end{tabular}
\end{table}

\section{Inversion of the Min-SNR FID Curve}
\label{app:inversion}

Table~\ref{tab:inversion} is computed as follows (script \texttt{analysis/matched\_threshold.py} in our code; seeds 0 and 1 agree to two decimals). Two kinds of uncertainty enter, and only the first is quantified. For sampling uncertainty, the three seeds of each configuration are treated as a normal sample with unknown mean and variance under the standard noninformative prior, so each configuration's mean FID has posterior $\bar F+(S_F/\sqrt3)\,\tau$, with $\bar F$ and $S_F$ the seed mean and standard deviation and $\tau$ Student $t$ on 2 degrees of freedom, independently across configurations; we take 400{,}000 joint draws. The rest are modeling assumptions. For each draw the Min-SNR curve is interpolated piecewise linearly in $\log\gamma$ through $\gamma\in\{2.5,5,10,20\}$, the range over which the observed means decrease, and $\gamma^\star$ is the threshold at which it equals the drawn FID $y$ of our weight. Every draw is handled by the same rule.
\begin{enumerate}
\item If $y$ is at least the drawn Min-SNR FID at $\gamma=2.5$, the draw is assigned $\gamma^\star<2.5$, whatever the shape of the rest of the curve.
\item Otherwise $\gamma^\star$ is the first (smallest-$\gamma$) crossing of the interpolated curve with level $y$. This settles draws whose curve crosses more than once, which happens in at most $4\%$ of draws.
\item If there is no crossing, $y$ lies below all four drawn Min-SNR values, and the draw is assigned $\gamma^\star>20$.
\end{enumerate}
Draws whose drawn curve is not monotone ($18.6\%$ of joint draws) go through the same rule. Quantiles that fall in a censored region are reported as bounds. At $\lambda=0.05$ the cases occur with probabilities $0.02$ (below $2.5$), $0.19$ (segment $[2.5,5]$), $0.32$ (segment $[5,10]$), $0.45$ (segment $[10,20]$) and $0.02$ (above $20$); at $\lambda=0.1$ with $0.08$, $0.43$, $0.27$, $0.18$ and $0.04$; and at $\lambda=0.2$ with $0.06$, $0.83$, $0.10$, $0.01$ and $0.002$.

\begin{table}[ht]
\centering
\small
\caption{Posterior for the FID-matching Min-SNR threshold $\gamma^\star(\lambda)$, obtained by inverting the Min-SNR FID curve over $\gamma\in\{2.5,5,10,20\}$ (procedure below). Draws assigned below $\gamma=2.5$ or above $\gamma=20$ are censored, so some quantiles are reported as bounds. The intervals reflect sampling uncertainty in the seed means, not the interpolation assumptions.}
\label{tab:inversion}
\begin{tabular}{lcccc}
\toprule
$\lambda$ & $\Pr(\gamma^\star<1/(2\lambda))$ & $\Pr(\gamma^\star<1/\lambda)$ & median of $\gamma^\star\lambda$ & 90\% interval for $\gamma^\star\lambda$ \\
\midrule
$0.05$ & $0.52$ & $0.98$ & $0.49$ & $(0.18,\ 0.77)$ \\
$0.1$  & $0.51$ & $0.78$ & $0.49$ & $(<0.25,\ 1.67)$ \\
$0.2$  & $0.06$ & $0.89$ & $0.70$ & $(<0.5,\ 1.26)$ \\
\bottomrule
\end{tabular}
\end{table}

Within these assumptions the handling rules matter little. Excluding the nonmonotone draws changes $\Pr(\gamma^\star<1/\lambda)$, the median and the $95$th percentile of $\gamma^\star\lambda$ to $0.98$, $0.50$ and $0.76$ at $\lambda=0.05$ (from $0.98$, $0.49$ and $0.77$), to $0.78$, $0.54$ and $1.65$ at $\lambda=0.1$ (from $0.78$, $0.49$ and $1.67$), and to $0.90$, $0.72$ and $1.17$ at $\lambda=0.2$ (from $0.89$, $0.70$ and $1.26$). Interpolating linearly in $\gamma$ instead of $\log\gamma$ leaves the probabilities unchanged, moves the medians to $0.49$, $0.50$ and $0.74$, and moves the $95$th percentiles to $0.81$, $1.74$ and $1.33$. Neither check addresses whether the true curve between grid points resembles the interpolation at all.

\end{document}